\documentclass[prd,aps,nofootinbib,notitlepage,preprintnumbers
]{revtex4}
\usepackage{graphicx,epsf,amsmat
h,amsfonts,amssymb,amsbsy}
\usepackage{epsfig}
\newcommand{\mat}{\left ( \begin{array}}
\newcommand{\emat}{\end{array} \right )}
\newcommand{\vect}{\left ( \begin{array}{c}}
\newcommand{\evect}{\end{array} \right )}

\newcommand{\Det}{\mathop{\rm Det}\nolimits}

\def\cP{\mathcal P}
\def\cT{\mathcal T}

\begin{document}


\title{Spontaneous non-Hermiticity in the (2+1)-dimensional Thirring model}

\author{M. M. Gubaeva$^{1}$, T. G. Khunjua $^{2}$, K. G. Klimenko $^{3}$, and R. N. Zhokhov $^{3,4}$ }

\affiliation{$^{1}$ Dubna State University (Protvino branch), 142281 Protvino, Moscow Region, Russia}
\affiliation{$^{2}$ The University of Georgia, GE-0171 Tbilisi, Georgia}
\affiliation{$^{3}$ State Research Center
of Russian Federation -- Institute for High Energy Physics,
NRC "Kurchatov Institute", 142281 Protvino, Moscow Region, Russia}
\affiliation{$^{4}$  Pushkov Institute of Terrestrial Magnetism, Ionosphere and Radiowave Propagation (IZMIRAN),
108840 Troitsk, Moscow, Russia}

\begin{abstract}
Using the Cornwall-Jackiw-Tomboulis effective action $\Gamma(S)$ for composite operators 
($S$ is the full fermion propagator), the phase structure of the massless (2 + 1)-dimensional Thirring model 
with four-component spinors is investigated in the Hartree-Fock (HF) approximation. In this case both $\Gamma(S)$ 
and its stationary (or HF) equation for the full fermion propagator $S$ are calculated in the first order of the bare 
coupling constant $G$. We have shown that there exist a well-defined dependence of $G\equiv G(\Lambda)$ on the cutoff 
parameter $\Lambda$ under which the HF equation is renormalized. In general, it has two sets, (i) and (ii), of 
solutions for fermion propagator corresponding to dynamical appearance of different mass terms in the model. In the 
case of set (i) the mass terms are Hermitian, but the solutions from the set (ii) correspond to a dynamical generation 
of the non-Hermitiam mass terms, i.e. to a spontaneous non-Hermiticity of the Thirring model. Despite this, the mass 
spectrum of the quasiparticle excitations of all non-Hermitian ground states is real. In addition, among these 
non-Hermitian phases there are both $\cP\cT$ symmetrical and non-symmetrical phases. Moreover, in contrast with 
previous investigations of this effect in other models, we have observed the spontaneous non-Hermiticity phenomenon 
also in the {\it massive} (2+1)-dimensional Thirring model.
\end{abstract}
\maketitle

\section{Introduction}

For a long time, it was believed that to describe quantum systems it is necessary to use theories 
with Hermitian Hamiltonians (or Lagrangians), since in this case the energy spectrum is real. 
However, in recent decades it has become clear that there are situations, especially in open physical systems 
interacting with the environment, that can be effectively considered in terms of non-Hermitian Hamiltonians.
In particular, non-Hermitian methods are used to describe processes in atomic, molecular and optical physics, 
mesoscopic and nuclear physics, to describe the properties of quark-gluon plasma and systems with dissipation, etc. 
(For more detailed description of the scientific areas in which non-Hermitian approaches are used, see, for example, 
in the recent papers \cite{Ashida,Kanazawa2}.) Moreover, it was argued that if non-Hermitian theories additionally 
have the spacetime reflection symmetry $\cP\cT$, then its energy spectrum is real \cite{Bender,Bender2}, 
i.e. the Hermiticity of the Hamiltonian is only a sufficient, but far from necessary, condition for the reality 
of the energy spectrum of the system. This assertion is confirmed in quantum mecanics and in scalar field theories, 
in which the non-Hermiticity together with the $\cP\cT$ symmetry leads to a real mass spectrum \cite{Bender3,Ai}.

In fact, non-Hermitian methods can also be used in the study of various fermionic systems, e.g., 
in condensed matter physics, as well as when considering the quark-gluon plasma when it is in the steady-stable
thermodynamically nonequilibrium regime in heavy-ion collisions, etc. And in these cases there are more opportunities 
for obtaining a real mass spectrum of quasiparticles. On the one hand, indeed, as the 
considerations of some (1 + 1)- and (3 + 1)-dimensional (D) and non-Hermitian field theory models 
with four-fermion interaction show, the $\cP\cT$-symmetry together with non-Hermiticity 
leads to a real spectrum of particle masses \cite{Klevansky,Felski,Felski2}. On the other hand, in the same 
paper \cite{Felski2} other non-Hermitian and anti-$\cP\cT$-symmetric extensions of the four-fermion 
models are also presented, in which, nevertheless, a real spectrum of fermion masses is also 
generated, i.e. in fact $\cP\cT$ symmetry of the model is not a necessary, but rather sufficient, condition for real 
fermion masses to exist. Thus, the relationship between the phenomena of non-Hermiticity, $\cP\cT$ symmetry 
and the reality of the energy spectrum in any quantum system remains a far from solved problem and 
deserves further study. Moreover, it should be noted one more feature of the non-Hermiticity 
phenomenon, which was observed recently just within the framework of the (3+1)-D Nambu--Jona-Lasinio (NJL)
\cite{NJL,Klevansky2,Buballa} and (2+1)-D Gross-Neveu (GN) \cite{GN} models with four-fermion interactions. Namely, in 
these {\it massless} models the non-Hermiticity can arise spontaneously \cite{Chernodub2,KKZ}. (Quite recently, 
it was noted in Ref. \cite{Mavromatos} that, perhaps, the phenomenon of spontaneous non-Hermiticity occurs 
also in some models of Yukawa type.) It means that (i) initial Lagrangians of the massless NJL and GN models are 
taken to be Hermitian and $\cP\cT$-symmetric, (ii) but, as it was proved in Refs. \cite{Chernodub2,KKZ},  
there exist non-Hermitian ground states of these systems. In other words, elementary excitations (with real masses)
over these ground states possess a non-Hermitian dynamics that can be described effectively in terms of 
non-Hermitian Lagrangians, in which non-Hermiticity arises spontaneously either due to Yukawa-type terms 
(as in the NJL model \cite{Chernodub2}) or due to non-Hermitian mass terms of the Fermi fields (as in the case of 
the (2+1)-D GN model \cite{KKZ}). It is also interesting to note that if initial Hermitian Lagrangian contains a
nonzero bare mass term, then the phenomenon of spontaneous non-Hermiticity is absent in the above-mentioned 
NJL and GN models.

In the present paper, we continue to study the features of the phenomenon of spontaneous non-Hermiticity,
but this time in the framework of the (2 + 1)-D Thirring model in which fermions are four-component spinors.
We show that, in contrast to the results obtained in Refs. \cite{Chernodub2,KKZ}, in the (2+1)-D Thirring model 
a more rich set of non-Hermitian phases can be generated spontaneously, both in the massless and massive variants 
of the model.

In this connection, it is necessary to note that over the past few decades much attention has been paid to the 
investigation of (2 + 1)-D field-theory models, which can be used to predict and study the condensed 
matter physics phenomena of planar nature such as quantum Hall effect, high-temperature 
superconductivity, low-energy graphene physics, etc. To a fairly large extent, these phenomena are 
usually considered in the framework of models with a four-fermion interaction. Among them is the Gross-Neveu model
\cite{Semenoff,Shovkovy,Gusynin,Mesterhazy,Vshivtsev,Khudyakov,Ebert,Kanazawa,Ramos,Rosenstein}, the Thirring model
\cite{Gomes,Hong,Hands,Itoh,Hyun,Krasnikov,Wipf,Wipf2,KGK1,KGK2,Janssen,Gies,Hands2} etc.
One of the reasons is that in the above mentioned (2+1)-D models the spontaneous symmetry breaking occurs dynamically, 
i.e. nonperturbatively and without taking into account additional scalar Higgs bosons. Moreover, despite the 
perturbative nonrenormalizability of these models, in the framework of nonperturbative 
approaches such as the large-$N$ technique, etc., they are renormalizable \cite{Rosenstein,Krasnikov}. 
And just using the $1/N$ approach, spontaneous symmetry breaking and the 
associated effect of dynamical generation of the fermion mass were investigated in these  
(2 + 1)-D models with four-fermion interaction. It should also be noted that other nonperturbative approaches, 
such as the optimized expansion technique \cite{Klimenko}, Gaussian variational methods \cite{Hyun,Kneur}, etc., 
predict qualitatively the same properties of the above-mentioned (2+1)-D models as the $1/N$ expansion. 

However, in the recent papers \cite{Tamaz1,KGK1,KGK2}, the so-called Hartree-Fock (HF) approach was used
in order to investigate the possibility of dynamical fermion mass generation in three-dimensional GN and 
Thirring models. \footnote{It is also worth mentioning  that the possibility of the dynamical appearance of the 
fermion mass in some non-Hermitian quantum field theory models has been investigated in Refs. 
\cite{Kanazawa2,Alexandre,Alexandre2,Mavromatos2}. Namely, in the first of these papers, the problem is considered 
within the framework of the $1/N$ expansion in the (3 + 1)-D NJL model (with a complex coupling constant), while in 
the remaining papers, for this purpose, the approach of the Dyson-Schwinger equation was used in non-Hermitian 
Yukawa-type models with additional four-fermion interaction term.}
The essence of the HF method consists, firstly, in the using of the Cornwall-Jackiw-Tomboulis (CJT) 
effective action for composite operators $\Gamma(S)$ \cite{CJT} in field theory models (here $S$ is the
full fermion propagator satisfying the stationary equation $\delta\Gamma/\delta S=0$), and, secondly, 
that $\Gamma(S)$ is considered in the first order in the coupling constant. As a result, in the framework of the 
models with four-fermion interaction, the stationary equation takes 
the form of the well-known Hartree--Fock equation for fermion mass operator \cite{Buballa,Klevansky}. 
Wherein, it turned out that in the region of large $N$ the HF method predicts qualitatively the same properties of 
the (2+1)-D GN and Thirring models as the leading order of the nonperturbative $1/N$-expansion method, widely used 
to study these models. 
But in the region of small $N$, where leading order of the $1/N$ approach is not applicable, the HF method predicts 
the existence of 
other nontrivial phases of the three-dimensional GN and Thirring models. In addition, the use of this research method 
made it possible to detect spontaneous non-Hermiticity in the (2 + 1)-D GN model \cite{KKZ}. Here we go further and 
study in the framework of the HF approach to (2+1)-D Thirring model the possibility of its spontaneous 
non-Hermiticity. Namely, we show that there can exists a non-Hermitian ground state 
of the model whose quasiparticle excitations have a real mass spectrum.

The paper is organized as follows. Section \ref{IIA} presents the $N$-flavor massless 
(2+1)-dimensional Thirring model constructed from four-component spinors. Here its invariance with respect to 
continuous $U(2N)$ transformations as well as under two spatial $\cP_k$ and two time $\cT_l$ reflections ($k,l=1,2$)
is established. The question of how different fermion-antifermion structures (possible massive terms of 
the model Lagrangian) are transformed under the influence of different $\cP_k\cT_l$ is clarified in it.
In section \ref{IIB} the CJT effective action $\Gamma(S)$ of the composite bilocal and bifermion operator 
$\overline\Psi (x)\Psi (y)$ is considered up to a first order in the bare coupling constant $G$ (it is the so-called 
Hartree-Fock approximation), which is actually the functional of the full 
fermionic propagator $S(x,y)$. In real situations, the propagator is a translation invariant 
solution of the corresponding stationary HF equation of the obtained CJT effective action. 
In section III  we show that for a some well-defined behavior of the coupling constant $G(\Lambda)$ vs $\Lambda$, 
there exists a set of different renormalized, i.e. without ultraviolet divergences, solutions of the HF equation for 
the propagator. Each of them corresponds to some phase in which a Hermitian fermion mass term arises dynamically. 
Section IV proves that HF equation has, in addition, a set of so-called non-Hermitian solutions. 
Each of them corresponds to a phase in which a non-Hermitian fermion mass term arises dynamically, and the 
massless Thirring model spontaneously becomes non-Hermitian. In this case, however, the mass spectrum of quasiparticle 
excitations of each of the ground states of these (non-Hermitian) phases is real. 
Finally, in the section V we show that the spontaneous non-Hermiticity of the (2+1)-D Thirring model can appear 
not only in the chiral limit, but also in the case when there is a nonzero (Hermitian) bare Haldane mass term.

\section{(2+1)-dimensional Thirring model and Hartree-Fock approach}

\subsection{Massless model, its symmetries, etc}\label{II.1}

The Lagrangian of the massless and $N$-flavored (2+1)-D Thirring model under 
consideration has the following form (see, e.g., in \cite{Itoh,Gies})
\begin{eqnarray}
 L=\overline \Psi_k\gamma^\nu i\partial_\nu \Psi_k&-& \frac {G}{2N}\left
(\overline \Psi_k\gamma^\mu\Psi_k\right )\left (\overline \Psi_l\gamma_\mu\Psi_l\right ),
\label{t1}
\end{eqnarray}
where for each $k=1,...,N$ the field $\Psi_k\equiv \Psi_k(t,x,y)$ is a (reducible) four-component Dirac spinor
(its spinor indices are omitted in Eq. (\ref{t1})), $\gamma^\nu$ 
($\nu=0,1,2$) are 4$\times$4 matrices acting in this four-dimensional spinor space (the algebra
of these $\gamma$-matrrices and their particular representation used in the present paper
is given in Appendix \ref{ApC}, where the matrices $\gamma^3,\gamma^5$ and $\tau=-i\gamma^3\gamma^5$
are also introduced), and the summation over repeated flavor $k,l$ and Lorentz $\mu,\nu$ indices is assumed in Eq. 
(\ref{t1}) and below. The bare coupling constant $G$ have a dimension of [mass]$^{-1}$. 

The set of all four-component spinor fields $\Psi_k$ ($k=1,...,N$) can be considered as a fundamental 
multiplet of the $U(N)$ flavor group, so the invariance of the 
Lagrangian (\ref{t1}) with respect to this group is obvious. However, it is not so obvious that in fact the 
continuous symmetry group of the (2+1)-D Thirring model (\ref{t1}) is wider and is $U(2N)$. This is easy to establish if we 
rewrite the expression (\ref{t1}) in terms of two-component spinors. Namely, for each fixed $k=1,...,N$ let us 
introduce the following structure of a four-component spinor,  
$\Psi_k^T=(\psi_{2k-1}^T,\psi_{2k}^T)$, where the symbol $T$ denotes the transposition operation, and $\psi_{2k-1}$ 
and $\psi_{2k}$ are two-component spinors. Then we have:
\begin{eqnarray}
 L_0\equiv \overline \Psi_k\gamma^\nu i\partial_\nu \Psi_k=\overline \psi_1\tilde\gamma^\nu i\partial_\nu \psi_1+
 \overline \psi_2\tilde\gamma^\nu i\partial_\nu \psi_2+\cdots +\overline \psi_{2N}\tilde\gamma^\nu i\partial_\nu
 \psi_{2N},\nonumber\\
 \overline \Psi_k\gamma^\nu\Psi_k=\overline \psi_1\tilde\gamma^\nu \psi_1+
 \overline \psi_2\tilde\gamma^\nu \psi_2+\cdots +\overline \psi_{2N}\tilde\gamma^\nu \psi_{2N},
\label{t2}
\end{eqnarray}
where $\tilde\gamma^\nu$ are $2\times 2$ matrices (see in Appendix \ref{ApC}). Assuming formally that the set of all
two-component spinors $\psi_{2k-1}$ and $\psi_{2k}$ ($k=1,..,N)$ is transformed by a fundamental representation of 
the $U(2N)$ group, it is easy to see that both the structures (\ref{t2}) and the entire Lagrangian (\ref{t1}) are 
invariant under this group. Notice that sometimes the continuous $U(2N)$ is called chiral symmetry group of the 
Thirring Lagrangian (\ref{t1}) \cite{Wipf}. The reason is that the $U(2N)$ group contains two chiral subgroups, 
$U(1)_{\gamma^5}$ and $U(1)_{\gamma^3}$, such that
\begin{eqnarray}
 U(1)_{\gamma^5}:&&\Psi_k\to  \exp(i\gamma^5\alpha)\Psi_k;~~
 \overline\Psi_k\to  \overline\Psi_k\exp(i\gamma^5\alpha),~~~~~k=1,...,N;\nonumber\\
 U(1)_{\gamma^3}:&&\Psi_k\to  \exp(i\gamma^3\tilde\alpha)\Psi_k;~~
 \overline\Psi_k\to  \overline\Psi_k\exp(i\gamma^3\tilde\alpha),~~~~~k=1,...,N.  \label{n4}
\end{eqnarray}

In addition, the Thirring Lagrangian (\ref{t1}) is also invariant under several discrete transformations
such as space parity $\cP$, time reversal $\cT$  and $\cP\cT$, whose action on four-component Fermi-fields 
$\Psi_k(t,x,y)$ in (2+1)-D spacetime should now be considered.

In (2+1) dimensions the space reflection, or parity, transformation $\cP$ is defined by
$(t,x,y)\stackrel{\mathcal P}{\longrightarrow} (t,-x,y)$. \footnote{In (2+1) spacetime dimensions, parity 
corresponds to inverting only one spatial axis \cite{Semenoff,Appelquist}, since the inversion 
of both axes is equivalent to rotating the entire space by $\pi$ angle.} To find the transformation of a four-component 
spinor field $\Psi_k (x)$ ($k=1,...,N$) under $\cP$, we postulate that the Lagrangian $L_0\equiv\overline
\Psi_k (x){\cal D}\Psi_k(x)$ of the free massless spinor fields $\Psi_k$, 
where ${\cal D}=i\gamma^0\partial_0+i\gamma^1\partial_1+i\gamma^2\partial_2$, 
remains intact under space reflection $\cP$. Hence (below, for the sake of breavity we denote by $x$ and $x'$ the 
set of coordinates $(t,x,y)$ and $(t,-x,y)$, respectively; moreover, to simplify the formulas, almost everywhere
in this section we omit the flavor index $k$ of the spinor fields, the summation over which, however, is implied 
in all bifermion structures) 
\begin{eqnarray}
L_0=\cP L_0 \cP^{-1}=\overline{\Psi^{\cP}}(x'){\cal D}'\Psi^{\cP}(x'), \label{tt4}
\end{eqnarray}
and
\begin{eqnarray}
{\cal D}'=\cP{\cal D}\cP^{-1}&=&i\gamma^0\partial_0-i\gamma^1\partial_1+i\gamma^2\partial_2,~~
\Psi^{\cP}(x')=\cP\Psi (x)\cP^{-1},~~\overline{\Psi^{\cP}} (x')=\cP\overline\Psi (x)\cP^{-1}. \label{nn4}
\end{eqnarray}
It follows from Eqs. (\ref{tt4}) and (\ref{nn4}) that $L_0$ is invariant under the 
action of parity operation only when $\cP$ is equal to one of the $\cP_5$ or $\cP_3$ transformations, where
\begin{eqnarray}
&&\cP_5:~ \Psi(x)\to\cP_5\Psi(x)\cP_5^{-1}\equiv\Psi^{\cP_5}(x')=  \gamma^5\gamma^1\Psi(x);~~
\overline\Psi(x)\to \cP_5\overline\Psi(x)\cP_5^{-1}\equiv\overline{\Psi^{\cP_5}} (x')= 
 \overline\Psi(x)\gamma^5\gamma^1,\nonumber\\
 &&\cP_3:~ \Psi(x)\to\cP_3\Psi(x)\cP_3^{-1}\equiv\Psi^{\cP_3}(x')=  \gamma^3\gamma^1\Psi(x);~~
\overline\Psi(x)\to \cP_3\overline\Psi(x)\cP_3^{-1}\equiv\overline{\Psi^{\cP_3}} (x')= 
 \overline\Psi(x)\gamma^3\gamma^1.\label{n2}
\end{eqnarray}
Moreover, it is easy to conclude that the Lagrangian (\ref{t1}) as a whole is also invariant under the action of 
each of the transformations $\cP_3$ and $\cP_5$. In addition, using Eq. (\ref{n2}) 
one can find how some Hermitian bispinors are transformed under the action of $\cP_5$,
\begin{eqnarray}
&&\overline\Psi(x)\Psi(x)\stackrel{\mathcal P_5}{\longrightarrow}\overline\Psi(x)\Psi(x);~~
\overline\Psi(x)i\gamma^5\Psi(x)\stackrel{\mathcal P_5}{\longrightarrow}-\overline\Psi(x)i\gamma^5
\Psi(x),\nonumber\\
&&\overline\Psi(x)\tau\Psi(x)\stackrel{\mathcal P_5}{\longrightarrow}-
\overline\Psi(x)\tau\Psi(x);~~\overline\Psi(x)i\gamma^3\Psi(x)\stackrel{\mathcal P_5}{\longrightarrow}\overline\Psi(x)i\gamma^3
\Psi(x).\label{nn3}
\end{eqnarray}
But the transformations of these bispinor combinations under the action of $\cP_3$ look like
\begin{eqnarray}
&&\overline\Psi(x)\Psi(x)\stackrel{\mathcal P_3}{\longrightarrow}\overline\Psi(x)\Psi(x);~~
\overline\Psi(x)i\gamma^5\Psi(x)\stackrel{\mathcal P_3}{\longrightarrow}\overline\Psi(x)i\gamma^5
\Psi(x),\nonumber\\
&&\overline\Psi(x)\tau\Psi(x)\stackrel{\mathcal P_3}{\longrightarrow}
\overline\Psi(x)\tau\Psi(x);~~\overline\Psi(x)i\gamma^3\Psi(x)\stackrel{\mathcal P_3}{\longrightarrow}
-\overline\Psi(x)i\gamma^3\Psi(x).\label{nnn3}
\end{eqnarray}

Now, let us consider the time reversal $\cT$ in the framework of the (2+1)-D Thirring model (\ref{t1}). 
In the (2+1)-dimensional spacetime it is defined as $(t,x,y)\stackrel{\mathcal T}{\longrightarrow} 
(-t,x,y)$. To determine how a four-component spinor field $\Psi$ is 
transformed under this operation in this case, we also assume from the very beginning (as in the case of
spatial reflection $\cP$) that the Lagrangian $L_0$ of free massless fermion
fields $\Psi$ remains invariant with respect to $\cT$, i.e. $L_0=\cT L_0 \cT^{-1}$, where (now, for the sake of 
breavity we denote in this case by $x$ and $x'$ the set of coordinates $(t,x,y)$ and $(-t,x,y)$, respectively) 
\begin{eqnarray}
\cT L_0 \cT^{-1}=\overline{\Psi^{\cT}} (x'){\cal D}'\Psi^{\cT}(x'),~~~
\Psi^{\cT}(x')=\cT\Psi (x)\cT^{-1},~~\overline{\Psi^{\cT}} (x')=\cT\overline\Psi (x)\cT^{-1},\label{n5}
\end{eqnarray}
and in this case ${\cal D}'=\cT {\cal D}\cT^{-1}$. In the following, it is very important to take 
into account that time-reversal operation $\cT$ (i) changes the sign of the time coordinate, $t\to -t$, 
and (ii) it is an anti-linear or anti-unitary one, which means that its action on any complex 
number or matrix $C$ transforms it into the complex conjugate $C^*$, i.e. $\cT C \cT^{-1}=C^*$ 
(for details, see, e.g., in Refs. \cite{Peskin2,Bender2,Chernodub}). Taking into account these 
(i) and (ii) properties of the $\cT$ transformation, we have
\begin{eqnarray}
{\cal D}'=i\gamma^{0*}\partial_0-i\gamma^{1*}\partial_1-i\gamma^{2*}\partial_2=
i\gamma^{0}\partial_0+i\gamma^{1}\partial_1-i\gamma^{2}\partial_2.\label{n6}
\end{eqnarray}
In the last equality we have used the relations $\gamma^{0*}=\gamma^{0}$, 
$\gamma^{1*}=-\gamma^{1}$ and $\gamma^{2*}=\gamma^{2}$ (see in Appendix \ref{ApC}). Now, it is 
rather evident from Eqs. (\ref{n5}) and (\ref{n6}) that $L_0$ is invariant under the 
action of time-reversal $\cT$ operation only when it is equal to one of the $\cT_5$ or $\cT_3$ transformations, where
\begin{eqnarray}
&&\cT_5:~\Psi(x)\to \cT_5\Psi(x)\cT_5^{-1}\equiv\Psi^{\cT_5}(x')=  
\gamma^5\gamma^2\Psi(x);~\overline\Psi(x)\to \cT_5\overline\Psi(x)\cT_5^{-1}\equiv\overline{\Psi^{\cT_5}} (x')= 
 \overline\Psi(x)\gamma^5\gamma^2,\nonumber\\
&&\cT_3:~\Psi(x)\to \cT_3\Psi(x)\cT_3^{-1}\equiv\Psi^{\cT_5}(x')=  
\gamma^3\gamma^2\Psi(x);~\overline\Psi(x)\to \cT_3\overline\Psi(x)\cT_3^{-1}\equiv\overline{\Psi^{\cT_3}} (x')= 
 \overline\Psi(x)\gamma^3\gamma^2.  \label{n7}
\end{eqnarray}
Using Eq. (\ref{n7}), it is easy to verify that not only $L_0$, but also the Lagrangian (\ref{t1}) are invariant
under both $\cT_5$ and $\cT_3$ transformations. Thus, there are two different time-inversion transformations 
$\cT_5$ and $\cT_3$, under which the massless (2+1)-D Thirring model is invariant. It is easy to obtain the following 
transformations of the above mentioned Hermitian bispinors under $\cT_5$,
\begin{eqnarray}
&&\overline\Psi(x)\Psi(x)\stackrel{\mathcal T_5}{\longrightarrow}\overline\Psi(x)\Psi(x);~~ 
\overline\Psi(x)i\gamma^3\Psi(x)\stackrel{\mathcal T_5}{\longrightarrow} -\overline\Psi(x)i\gamma^3\Psi(x); 
\nonumber\\
&&\overline\Psi(x)i\gamma^5\Psi(x)\stackrel{\mathcal T_5}{\longrightarrow}-\overline\Psi(x)i\gamma^5
\Psi(x);~~\overline\Psi(x)\tau\Psi(x)\stackrel{\mathcal T_5}{\longrightarrow}-\overline\Psi(x)\tau\Psi(x).\label{n8}
\end{eqnarray}
But under $\cT_3$ they are transformed as
\begin{eqnarray}
&&\overline\Psi(x)\Psi(x)\stackrel{\mathcal T_3}{\longrightarrow}\overline\Psi(x)\Psi(x);~~ 
\overline\Psi(x)i\gamma^3\Psi(x)\stackrel{\mathcal T_3}{\longrightarrow} \overline\Psi(x)i\gamma^3\Psi(x); 
\nonumber\\
&&\overline\Psi(x)i\gamma^5\Psi(x)\stackrel{\mathcal T_3}{\longrightarrow}\overline\Psi(x)i\gamma^5
\Psi(x);~~\overline\Psi(x)\tau\Psi(x)\stackrel{\mathcal T_3}{\longrightarrow}-\overline\Psi(x)\tau\Psi(x).\label{nn8}
\end{eqnarray}
Since the Thirring Lagrangian (\ref{t1}) is invariant with respect to each of the transformations $\cP_5$, $\cP_3$, 
$\cT_5$ and $\cT_3$, there exist four $\cP\cT$ transformations under which the model (\ref{t1}) is also invariant. They
are $\cP_5\cT_5$, $\cP_5\cT_3$, $\cP_3\cT_5$ and $\cP_3\cT_3$. Using the relations (\ref{nn3}), (\ref{nnn3}) and 
(\ref{n8}), (\ref{nn8}), one can easily find the transformations of the above mentioned Hermitian bispinors
under these $\cP_k\cT_l$ ($k,l=3,5$) transformations (see first four rows in Table I). In addition, the same Table I 
shows the action of all descrete $\cP_k\cT_l$ operators on some anti-Hermitian bispinors (under the action of the 
Hermitian conjugation, they change their sign to the opposite). 

Due to the  symmetry of the model (\ref{t1}) with respect to the above mentioned continuous and discrete 
transformations, different mass 
terms are prohibited to appear in this Lagrangian perturbatively. Indeed, the most popular mass term has the 
form $m_D\overline \Psi_k \Psi_k$ (for it we use the notation Dirac mass term), but it breaks, as it follows from Eq. 
(\ref{n4}), both $U(1)_{\gamma^5}$ and $U(1)_{\gamma^3}$ chiral symmetries of the model, although this mass term 
is $\cP_k$ and $\cT_l$ ($k,l=3,5$) even. There is another well-known 
expression for fermion mass that is often discussed in the literature. This is a mass term of 
the form $m_H\overline \Psi_k\tau \Psi_k$ (recall, here the 4$\times$4 matrix $\tau$ is defined 
in Appendix \ref{ApC}) and sometimes it is refered to as the Haldane mass term (see, e.g., in Refs. 
\cite{Ebert}). \footnote{The appearance of the Haldane mass term is related 
to the parity anomaly in (2+1) dimensions, to generation of the Chern-Simons topological mass 
of gauge fields \cite{Gomes2,Klimenko2}, as well as to the integer quantum Hall effect in planar 
condensed matter systems without external magnetic field, etc \cite{Haldane}.} 
But the nonzero Haldane mass term is not invariant under the parity $\cP_5$, although it is 
invariant under chiral symmetries (\ref{n4}) and $\cP_3$. There exist two another, chiral, mass 
terms, $i m_3\overline \Psi_k \gamma^3\Psi_k$ and $i m_5\overline \Psi_k \gamma^5\Psi_k$, the 
dynamical generation of which we are also going to study here. The first one 
breaks $U(1)_{\gamma^3}$ symmetry (but it is invariant under $U(1)_{\gamma^5}$ and $\cP_5$), whereas the second 
mass term is not invariant under $U(1)_{\gamma^5}$ and $\cP_5$ transformations (but it is $U(1)_{\gamma^3}$ and $\cP_3$
symmetrical). So the Dirac $m_D$, Haldane $m_H$ and $m_3,m_5$ masses can not appear perturbatively in the chirally 
$U(1)_{\gamma^5}$, $U(1)_{\gamma^3}$, parity $\cP_5$, $\cP_3$, etc. invariant massless Lagrangian (\ref{t1}). 

Nevertheless, as can be seen from the analysis of the phase structure 
of the model (\ref{t1}) carried out within HF approach 
(see in the text below), all the above-mentioned masses may be dynamically, i.e. nonperturbatively, induced in the 
(2+1)-D Thirring model. This 
means that there exists such a behavior of the bare coupling constant $G\equiv 
G(\Lambda)$ vs the cutoff regularization parameter $\Lambda$ that a phase with one or another finite 
nonzero fermion mass can be dynamically generated in the model. 

\renewcommand{\arraystretch}{1.9}
\renewcommand{\tabcolsep}{1.0cm}
\begin{table*}[!t]
\centering
\begin{tabular}{|c | c c c c|}\hline\hline
bispinors$\big\backslash$transformations   & $\cP_5\cT_5$    &  $\cP_5\cT_3$   &   $\cP_3\cT_5$   &  $\cP_3\cT_3$   \\ \hline
$\overline\Psi(x)\Psi(x)$   &    even   &   even   &   even   &  even  \\
$\overline\Psi(x)i\gamma^5\Psi(x)$  &   even   &  odd    &   odd   &  even  \\
$\overline\Psi(x)i\gamma^3\Psi(x)$  &   odd   &  even    &   even   &  odd  \\
$\overline\Psi(x)\tau\Psi(x)$  & even  &  even   &   odd   &  odd  \\ \hline
$i\overline\Psi(x)\Psi(x)$   &    odd   &   odd   &   odd   &  odd  \\
$\overline\Psi(x)\gamma^5\Psi(x)$  &   odd   &  even   &   even   &  odd   \\
$\overline\Psi(x)\gamma^3\Psi(x)$  &   even   &  odd    &   odd   &  even  \\
$\overline\Psi(x)i\tau\Psi(x)$  & odd  &  odd   &   even   &  even  \\
 \hline\hline
\end{tabular}
\caption{\label{Tab1} Behavior of various Hermitian (from rows 1 through 4) and anti-Hermitian 
(from rows 5 through 8) bispinor structures under different $\cP_k\cT_l$ 
transformations ($k,l=3,5$). Here ''even`` means that a bispinor remains intact, ''odd`` -- that a bispinor changes 
sign to the opposite under the action of $\cP_k\cT_l$.}
\end{table*}\label{IIA}

\subsection{Hartree-Fock approach}\label{IIB}

Let us define $Z(K)$, the generating functional of the Green's functions of bilocal 
fermion-antifermion composite operators $\sum_{k=1}^N\overline\Psi_k^\alpha(x)\Psi_{k\beta}(y)$ 
in the framework of a (2+1)-D Thirring model (\ref{t1}) (the corresponding technique 
for theories with four-fermion interaction is elaborated in details, e.g., in Ref. \cite{Rochev}) 
\begin{eqnarray}
 Z(K)\equiv\exp(iNW(K))=\int {\cal D}\overline\Psi_k {\cal D}\Psi_k \exp\Big
(i\Big [ I(\overline\Psi,\Psi)+\int d^3xd^3y\overline\Psi_k^\alpha(x)K_\alpha^\beta(x,y)
\Psi_{k\beta}(y) \Big ]\Big ), \label{36}
\end{eqnarray}
where $\alpha,\beta =1,2,3,4$ are spinor indices, $K_\alpha^\beta(x,y)$ is a bilocal source of 
the fermion bilinear composite field $\overline\Psi_k^\alpha(x)\Psi_{k\beta}(y)$ (recall that in 
all expressions the summation over repeated indices is assumed). 
\footnote{We denote a matrix element of an arbitrary matrix (operator) $\hat A$ acting in the 
four dimensional spinor space by the symbol $A^\alpha_\beta$, where the upper (low) index 
$\alpha $($\beta$) is the column (row) number of the matrix $\hat A$. In particular, the matrix 
elements of any $\gamma^\mu$ matrix is denoted by $(\gamma^\mu)^\alpha_\beta$. }
Moreover, $I(\overline\Psi,\Psi)=\int Ld^3x$, where $L$ is the Lagrangian (\ref{t1}) of the  
(2+1)-dimensional Thirring model under consideration. Hence,
\begin{eqnarray}
&&I(\overline\Psi ,\Psi)=\int d^3xd^3y\overline\Psi_k^\alpha(x)D_\alpha^\beta(x,y)\Psi_{k\beta} (y)+
I_{int}(\overline\Psi_k^\alpha\Psi_{k\beta}),~~D_\alpha^\beta(x,y)=
\left(\gamma^\nu\right)_\alpha^\beta i\partial_\nu\delta^3(x-y),\nonumber\\
&&I_{int}= -\frac {G}{2N}\int d^3x\left (\overline \Psi_k\gamma^\mu\Psi_k\right )
\left (\overline \Psi_l\gamma_\mu\Psi_l\right )\nonumber\\
&&=-\frac {G}{2N}\int d^3xd^3td^3ud^3v\delta^3 (x-t)\delta^3 (t-u)\delta^3 (u-v) \overline
\Psi_k^\alpha(x)(\gamma^\mu)_\alpha^\beta\Psi_{k\beta}(t) \overline\Psi_l^\rho(u)
(\gamma_\mu)_\rho^\xi\Psi_{l\xi}(v). \label{360}
\end{eqnarray}
Note that in Eq. (\ref{360}) and similar expressions below, $\delta^3(x-y)$ denotes the 
three-dimensional Dirac delta function. There is an alternative expression for $Z(K)=\exp(iNW(K))$,
\begin{eqnarray}
\exp(iNW(K))&=&\exp\Big (iI_{int}\Big (-i\frac{\delta}{\delta K}\Big )\Big )\int 
{\cal D}\overline\Psi_k {\cal D}\Psi_k \exp\Big (i
\int d^3xd^3y\overline\Psi_k(x)\Big [D(x,y)+K(x,y)\Big ]\Psi_k (y)\Big )
\nonumber\\&=&\exp\Big (iI_{int}\Big (-i\frac{\delta}{\delta K}\Big )\Big )\Big 
[\det\big (D(x,y)+K(x,y)\big )\Big ]^N\nonumber\\&=&\exp\Big (iI_{int}\Big 
(-i\frac{\delta}{\delta K}\Big )\Big )\exp \Big [N{\rm Tr}\ln \big (D(x,y)+K(x,y)\big )\Big ],
\label{036}
\end{eqnarray}
where instead of each bilinear form $\overline\Psi_k^\alpha(s)\Psi_{k\beta}(t)$ appearing in $I_{int}$ of
the Eq. (\ref{360}) we use a variational derivative $-i\delta /\delta K^\beta_\alpha (s,t)$.
Moreover, the Tr-operation in Eq. (\ref{036}) means the trace both over spacetime and spinor coordinates. 
The effective action (or CJT effective action) of the composite bilocal and bispinor operator 
$\overline\Psi_k^\alpha(x)\Psi_{k\beta}(y)$ is defined as a functional $\Gamma (S)$ of the full 
fermion propagator $S^\alpha_\beta(x,y)$ by a Legendre transformation of the functional 
$W(K)$ entering in Eqs. (\ref{36}) and (\ref{036}),
\begin{eqnarray}
\Gamma (S)=W(K)-\int d^3xd^3y S^\alpha_\beta(x,y)K_\alpha^\beta(y,x),
\label{0360}
\end{eqnarray}
where
\begin{eqnarray}
S^\alpha_\beta(x,y)=\frac{\delta W(K)}{\delta K_\alpha^\beta(y,x)}.
 \label{37}
\end{eqnarray}
Taking into account the relation (\ref{36}), it is clear that at $K(x,y)=0$ the quantity $S(x,y)$ is the full fermion 
propagator of the Thirring model (\ref{t1}). Hence, in order to construct the CJT effective action $\Gamma (S)$ of Eq. 
(\ref{0360}), it is necessary to solve Eq. (\ref{37}) with respect to $K$ and then to use the obtained expression for 
$K$ (it is a functional of $S$) in Eq. (\ref{0360}). It follows from the definitions (\ref{0360})-(\ref{37}) that
\begin{eqnarray}
\frac{\delta\Gamma (S)}{\delta S^\alpha_\beta(x,y)}=\int d^3ud^3v\frac{\delta W(K)}{\delta K^\mu_\nu(u,v)}
\frac{\delta K^\mu_\nu(u,v)}{\delta S^\alpha_\beta(x,y)}-K_\alpha^\beta(y,x)-\int d^3ud^3v S_\mu^\nu(v,u)
\frac{\delta K^\mu_\nu(u,v)}{\delta S^\alpha_\beta(x,y)}. \label{037}
\end{eqnarray}
(In Eq. (\ref{037}) and below, the Greek letters $\alpha,\beta,\mu,\nu,$ etc, also denote
the spinor indices, i.e. $\alpha,...\nu,...=1,...,4$.) Now, due to the relation (\ref{37}), it is easy to see 
that the first term in Eq. (\ref{037}) cansels there the last term, so
\begin{eqnarray}
\frac{\delta\Gamma (S)}{\delta S^\alpha_\beta(x,y)}=-K_\alpha^\beta(y,x).
\label{370}
\end{eqnarray}
Hence, in the true Thirring model (\ref{t1}), in which bilocal sources $K_\alpha^\beta(y,x)$ are zero, the full 
fermion propagator is a solution of the following stationary equation,
\begin{eqnarray}
\frac{\delta\Gamma (S)}{\delta S^\alpha_\beta(x,y)}=0.
\label{0370}
\end{eqnarray}
Note that in the nonperturbative CJT approach the stationary/gap equation (\ref{0370}) 
for fermion propagator $S^\beta_\alpha(x,y)$  is indeed a Schwinger--Dyson equation \cite{Rochev}.
Further, in order to simplify the calculations and obtain specific information about 
the phase structure of the model, we calculate the effective action (\ref{0360}) up to a 
first order in the coupling $G$. 

In the literature, such an approach to effective action $\Gamma (S)$ of any model, 
including field theories with four-fermion interaction, is usually called the Hartree-Fock (HF) approximation 
\cite{CJT,KGK1,KGK2} (a more detailed justification for this name is given at the end of this section).

In the first order in coupling constant $G$, we have (detailed calculations are given in Appendix B of the 
paper \cite{KGK2})
\begin{eqnarray}
&&\Gamma (S)=-i{\rm Tr}\ln \big (-iS^{-1}\big )+\int d^3xd^3y S^\alpha_\beta(x,y)D_\alpha^\beta(y,x)
\nonumber\\
&&-\frac{G}2\int d^3x~ {\rm tr}\big [\gamma^\rho S(x,x)
\big ] {\rm tr}\big [\gamma_\rho S(x,x)\big ]
+\frac{G}{2N}\int d^3x~ {\rm tr}\Big [\gamma^\rho S(x,x)\gamma_\rho S(x,x)\Big ].
 \label{n420}
\end{eqnarray}
Notice that in Eq. (\ref{n420}) the symbol tr means the trace of an operator 
over spinor indices only, but Tr is the trace both over spacetime coordinates and spinor 
indices. Moreover, there the operator $D(x,y)$ is introduced in Eq. (\ref{360}). The stationary, or Schwinger--Dyson, 
equation (\ref{0370}) for the CJT effective action (\ref{n420}) looks like 
\begin{eqnarray}
-i\Big [S^{-1}\Big ]^\beta_\alpha(x,y)-D_\alpha^\beta(x,y)&=&-G\big (\gamma^\rho\big )^\beta_\alpha{\rm tr}
\big [\gamma_\rho S(x,y)\big ]\delta^3(x-y)
+\frac{G}N \big [\gamma^\rho S(x,y)\gamma_\rho\big ]^\beta_\alpha\delta^3 (x-y).
\label{n0420}
\end{eqnarray}
Now suppose that $S(x,y)$ is a translationary invariant operator. Then 
\begin{eqnarray}
S^\beta_\alpha(x,y)\equiv S^\beta_\alpha(z)&=&\int\frac{d^3p}{(2\pi)^3}
\overline{S^\beta_\alpha}(p)e^{-ipz},~~~\overline{S^\beta_\alpha}(p)=\int d^3z S^\beta_\alpha(z)e^{ipz},\nonumber\\
\Big (S^{-1}\Big )^\beta_\alpha(x,y)&\equiv& \Big (S^{-1}\Big )^\beta_\alpha(z)=\int\frac{d^3p}{(2\pi)^3}
\overline{(S^{-1})^\beta_\alpha}(p)e^{-ipz},\label{n43}
\end{eqnarray}
where $z=x-y$ and $\overline{S^\beta_\alpha}(p)$ is a Fourier transformation of $ S^\beta_\alpha(z)$.
After Fourier transformation, Eq. (\ref{n0420}) takes the form
\begin{eqnarray}
-i\overline{(S^{-1})^\beta_\alpha}(p)- (\hat p)^\beta_\alpha&=&
-G\big (\gamma^\rho\big )^\beta_\alpha\int\frac{d^3q}{(2\pi)^3}~{\rm tr}
\big [\gamma_\rho\overline{S}(q)\big ]
+\frac{G}N \int\frac{d^3q}{(2\pi)^3}~\big [\gamma^\rho\overline{S}(q)\gamma_\rho\big ]^\beta_\alpha,
 \label{n043}
\end{eqnarray}
where $\hat p=p_\nu\gamma^\nu$. It is clear from Eq. (\ref{n043}) that in the framework of the four-fermion model 
(\ref{t1}) 
the Schwinger-Dyson equation for fermion propagator $\overline{S}(p)$ reads in the first order
in $G$ like the Hartree-Fock equation for its self-energy operator $\Sigma(p)$ (the last quantity is nothing 
but the expression on the left side of this equation). In particular, 
the first term on the right-hand side of Eq. (\ref{n043}) is the so-called Hartree contribution, whereas the last 
term there is the Fock contribution to the fermion self energy (for details, see, e.g., the section 4.3.1 in Ref. 
\cite{Buballa} or the section II C in Ref. \cite{Klevansky}). Also for this reason, the stationary equation (\ref{n043}) will be called the HF equation.

Finally note that both the CJT (or HF) effective action (\ref{n420}) and its stationary HF equation 
(\ref{n0420})-(\ref{n043}), in which $G$ is a bare coupling constant, contain ultraviolet (UV)
divergences and need to be renormalized. In the next sections, using a rather general ansatz for fermion
propagator $\overline{S}(p)$, we find the corresponding mode of the coupling constant $G$ 
behavior vs cutoff parameter $\Lambda$, such that  there occurs a renormalization of the gap HF equation 
(\ref{n043}), and it is possible to obtain its finite solution in the limit $\Lambda\to\infty$.

\section{Possibility for dynamical generation of Hermitian mass terms }\label{III}

In the present section, we study in the HF approximation the possibility of the dynamic appearance of 
the Hermitian mass term ${\cal M}_H$ in the model (\ref{t1}). It has the form
\begin{eqnarray}
{\cal M}_H=\overline\Psi_k (m_H\tau+m_D+im_5\gamma^5+im_3\gamma^3)\Psi_k. \label{t13}
\end{eqnarray}
It means that we should find such a solution $\overline S(p)$ of the 
stationary HF equation (\ref{n043}) that 
\begin{eqnarray}
\overline{S^{-1}}(p)=i\left (\hat p+m_H\tau+m_D+im_5\gamma^5+im_3\gamma^3\right ),\label{t140}
\end{eqnarray}
where $m_D$, $m_H$, $m_3$ and $m_5$ are finite unknown real quantities. Using any program of analytical calculations,
it is easy to obtain the propagator $\overline S(p)$ which is indeed a matrix inverse to the 4$\times$4 matrix
(\ref{t140}),
\begin{eqnarray}
&&\overline {S}(p)= 
\frac{-i}{\det(p)}  \left (\begin{array}{cc}
a(p)I~, & b(p)I\\
\overline b(p)I~, &\overline a(p)I
\end{array}\right )+\frac{i}{\det(p)}  \left (\begin{array}{cc}
[(\Sigma-m_H)^2-p^2]\tilde p~, & 2m_H(m_5+im_3)\tilde p\\
-2m_H(m_5-im_3)\tilde p~, &-[(\Sigma-m_H)^2-p^2]\tilde p
\end{array}\right ), \label{t14}
\end{eqnarray}
where $I$ is a unit 2$\times$2 matrix, while $\tilde p\equiv p_\nu\tilde\gamma^\nu$ is also 2$\times$2 matrix (the 
corresponding $\tilde\gamma$-matrices are defined in Appendix A), and $p^2=p_0^2-p_1^2-p_2^2$. Moreover, we use in 
Eq. (\ref{t14}) the following notations
\begin{eqnarray}
a(p)&=&m_D (\Sigma^2-m_H^2-p^2)-m_H(\Sigma^2-m_H^2+p^2);~~b(p)=-(m_5+im_3)(\Sigma^2-m_H^2-p^2),\nonumber\\
\bar a(p)&=&m_D (\Sigma^2-m_H^2-p^2)+m_H(\Sigma^2-m_H^2+p^2),~~\bar b(p)=-(-m_5+im_3)(\Sigma^2-m_H^2-p^2),
\nonumber\\
&&{\rm det(p)}=\left (p^2-(m_H+\Sigma)^2\right )
\left (p^2-(m_H-\Sigma)^2\right ),~~\Sigma=\sqrt{m_D^2+m_3^2+m_5^2}.
\label{t16}
\end{eqnarray}
(Notice that $\det(p)$ is indeed a determinant of the 4$\times$4 matrix (\ref{t140}).)
We would like to emphasize that the first term on the right-hand side of the equality (\ref{t14}) is an even vs
each momentum $p^\mu$. In contrast, the second term there is odd with respect to each momentum 
$p^\mu$. Therefore, after integration over the momenta, it will not contribute at all to the right side of the HF equation (\ref{n043}) (however, its 
contribution to the expression (\ref{n420}) for the CJT effective action in the HF approximation is nonzero). 
Taking into account this circumstance, after 
substituting the (\ref{t140})-(\ref{t16}) expressions into the HF equation (\ref{n043}), we obtain 
for the quantities $m_D$, $m_H$, $m_3$ and $m_5$ the following {\it unrenormalized} system of gap equations
\begin{eqnarray}
m_H&=&\frac{3iG}{2N}\int\frac{d^3p}{(2\pi)^3}\left\{
\frac{\Sigma+m_H}{p^2-(\Sigma+m_H)^2}-\frac{\Sigma-m_H}{p^2-(\Sigma-m_H)^2}\right\},\nonumber\\
m_D&=&\frac{3iG}{2N}\frac{m_D}{\Sigma}\int\frac{d^3p}{(2\pi)^3}\left\{
\frac{\Sigma+m_H}{p^2-(\Sigma+m_H)^2}+\frac{\Sigma-m_H}{p^2-(\Sigma-m_H)^2}\right\},\nonumber\\
m_5&=&\frac{3iG}{2N}\frac{m_5}{\Sigma}\int\frac{d^3p}{(2\pi)^3}\left\{
\frac{\Sigma+m_H}{p^2-(\Sigma+m_H)^2}+\frac{\Sigma-m_H}{p^2-(\Sigma-m_H)^2}\right\},\nonumber\\
m_3&=&\frac{3iG}{2N}\frac{m_3}{\Sigma}\int\frac{d^3p}{(2\pi)^3}\left\{
\frac{\Sigma+m_H}{p^2-(\Sigma+m_H)^2}+\frac{\Sigma-m_H}{p^2-(\Sigma-m_H)^2}\right\},
\label{t18}
\end{eqnarray}
Note that three-dimensional integrals in Eq. (\ref{t18}) are UV divergent and must
be regularized. Performing in these integrals a Wick rotation, $p_0\to i p_3$, and then using in the obtained 
three-dimensional Euclidean integration space the spherical coordinate system, $p_3=p\cos\theta, 
p_1=p\sin\theta\cos\phi, p_2=p\sin\theta\sin\phi$, we have (after integration over angles, $0\le\theta\le\pi, 
0\le\phi\le2\pi$, and cutting off the region of integration of the variable $p$, $0\le p\le \Lambda$) 
the following {\it regularized} gap system
\begin{eqnarray}
m_H&=&\frac{3G}{2N}\int_0^\Lambda\frac{p^2dp}{2\pi^2}\left\{
\frac{\Sigma+m_H}{p^2+(\Sigma+m_H)^2}-\frac{\Sigma-m_H}{p^2+(\Sigma-m_H)^2}\right\},\nonumber\\
m_D&=&\frac{3G}{2N}\frac{m_D}{\Sigma}\int_0^\Lambda\frac{p^2dp}{2\pi^2}\left\{
\frac{\Sigma+m_H}{p^2+(\Sigma+m_H)^2}+\frac{\Sigma-m_H}{p^2+(\Sigma-m_H)^2}\right\},\nonumber\\
m_5&=&\frac{3G}{2N}\frac{m_5}{\Sigma}\int_0^\Lambda\frac{p^2dp}{2\pi^2}\left\{
\frac{\Sigma+m_H}{p^2+(\Sigma+m_H)^2}+\frac{\Sigma-m_H}{p^2+(\Sigma-m_H)^2}\right\},\nonumber\\
m_3&=&\frac{3G}{2N}\frac{m_3}{\Sigma}\int_0^\Lambda\frac{p^2dp}{2\pi^2}\left\{
\frac{\Sigma+m_H}{p^2+(\Sigma+m_H)^2}+\frac{\Sigma-m_H}{p^2+(\Sigma-m_H)^2}\right\},
\label{t19}
\end{eqnarray}
where $\Lambda$ is a cutoff parameter. Since
\begin{eqnarray}
\int_0^\Lambda\frac{p^2}{p^2+M^2}dp=\Lambda-\frac \pi 2 |M|+
M{\cal O}\left (\frac M\Lambda\right ), \label{t6}
\end{eqnarray}
the system of gap regularized equations (\ref{t19}) can be reduced to the following form 
\begin{eqnarray}
\frac{m_H}{A}&=&2m_H\Lambda-\frac{\pi}{2}\Big[ (\Sigma+m_H)|\Sigma+m_H|-(\Sigma-m_H)|\Sigma-m_H|\Big ]
+m_H{\cal O}\left (\frac{m_H}\Lambda\right ),\nonumber\\
\frac{m_D}{A}&=&2m_D\Lambda-\frac{\pi}{2}\frac{m_D}{\Sigma}\Big[ (\Sigma+m_H)|\Sigma+m_H|+(\Sigma-m_H)|\Sigma-m_H|\Big ]
+m_D{\cal O}\left (\frac{m_D}\Lambda\right ),
\nonumber\\
\frac{m_3}{A}&=&2m_3\Lambda-\frac{\pi}{2}\frac{m_3}{\Sigma}\Big[ (\Sigma+m_H)|\Sigma+m_H|+(\Sigma-m_H)|\Sigma-m_H|\Big ]
+m_3{\cal O}\left (\frac{m_3}\Lambda\right ),
\nonumber\\
\frac{m_5}{A}&=&2m_5\Lambda-\frac{\pi}{2}\frac{m_5}{\Sigma}\Big[ (\Sigma+m_H)|\Sigma+m_H|+(\Sigma-m_H)|\Sigma-m_H|\Big ]
+m_5{\cal O}\left (\frac{m_5}\Lambda\right ),
\label{t71}
\end{eqnarray}
where $A=\frac{3G}{4N\pi^2}$. To remove the UV divergences from Eqs. (\ref{t71}), we suppose that the bare quantity
$A\equiv A(\Lambda)$ has the following asymptotic behavior at $\Lambda\to\infty$
\begin{eqnarray}
\frac{1}{A(\Lambda)}&=&2\Lambda+\frac{\pi}{2}g+g{\cal O}\left (\frac{g}\Lambda\right ),
\label{t73}
\end{eqnarray}
where $g$ is a finite $\Lambda$-independent and renormalization group invariant parameter with dimension of [mass]. 
It is clear from Eq. (\ref{t73}) that at sufficiently high values of $\Lambda$
\begin{eqnarray}
G\equiv G(\Lambda)&=&\frac{2\pi^2N}{3\Lambda}-\frac{\pi^3Ng}{6\Lambda^2}+\cdots.
\label{t75}
\end{eqnarray}
In this case, at $\Lambda\to\infty$ the system of stationary HF equations (\ref{t71}) takes the following 
{\it renormalized } form
\begin{eqnarray}
&&gm_H+ (\Sigma+m_H)|\Sigma+m_H|-(\Sigma-m_H)|\Sigma-m_H|=0,\nonumber\\
&&m_D\Big\{g+ \frac 1\Sigma\Big[(\Sigma+m_H)|\Sigma+m_H|+(\Sigma-m_H)|\Sigma-m_H|\Big]\Big\}=0,\nonumber\\
&&m_3\Big\{g+ \frac 1\Sigma\Big[(\Sigma+m_H)|\Sigma+m_H|+(\Sigma-m_H)|\Sigma-m_H|\Big]\Big\}=0,\nonumber\\
&&m_5\Big\{g+ \frac 1\Sigma\Big[(\Sigma+m_H)|\Sigma+m_H|+(\Sigma-m_H)|\Sigma-m_H|\Big]\Big\}=0.
\label{t74}
\end{eqnarray}
In the most general case, the system of gap HF equations (\ref{t74}) has several solutions.
To determine which of them is the most preferable, it is necessary to attract the so-called 
CJT effective potential $V(S)$ (in statistical physics, this quantity is called free energy), which is 
constructed on the basis of the CJT effective action by the following relation \cite{CJT}
\begin{eqnarray}
V(S)\int d^3x\equiv -\Gamma(S)\Big |_{\rm transl.-inv.~S(x,y) }. \label{t76}
\end{eqnarray}
To find CJT effective potential $V(S)$ in the Hartree-Fock approximation, we should use in Eq. (\ref{t76}) the 
expressions (\ref{n420}) and (\ref{t14}) for CJT effective action $\Gamma(S)$ and for the full fermion propagator 
$S(x,y)$, respectively. But in this case the obtained expression for $V(S)$ contains UV-divergences. However,
they are eliminated if bare coupling $G$ is constrained by the relation (\ref{t75}).  As a result, in the HF 
approach we have for the CJT effective potential $V(S)\equiv V(m_H,m_D,m_3,m_5)$ the following {\it renormalized} expression
(more detailed calculations are presented in Appendix \ref{ApB})
\begin{eqnarray}
V(m_H,m_D,m_3,m_5)\equiv V(m_H,\Sigma)=\frac{1}{12\pi}\Big (3 g\Sigma^2+3 g m_H^2 +2|\Sigma+m_H|^3+
2 |\Sigma-m_H|^3\Big ),\label{t77}
\end{eqnarray}
where $\Sigma=\sqrt{m_D^2+m_3^2+m_5^2}\ge 0$, and $g$ is some finite and renormalization group 
invariant quantity defined by Eq. (\ref{t73}) 
(notice that the expression (\ref{t77}) is valid up to unessential mass-independent infinite constant).
Note in addition that the gap equations (\ref{t74}) are also the stationary equations for the CJT effective 
potential (\ref{t77}). The global minimum point (GMP) of the function 
$V(m_H,m_D,m_3,m_5)$ determines the values of fermion masses $m_H,m_D,m_3,m_5$ which are generated dynamically in 
the massless Thirring model when coupling constant $G$ is constrained by the condition (\ref{t75}).

Let us consider the GMP of the function $V(m_H,\Sigma)$ (\ref{t77}) and its behavior vs $g$. First, 
note that this function is symmetric under the transformation $m_H\to -m_H$. 
So, for simplicity, it is sufficient to look for its GMP only in the region $\Sigma,m_H\ge 0$. Second, it is evident 
that at $g\ge 0$ the GMP of $V(m_H,\Sigma)$ lies at the point $(m_H=0,\Sigma=0)$, which means that no fermion 
mass terms are generated in this region, and all model symmetries discussed in Section \ref{II.1} remain intact. 

In contrast, if $g <0$ then effective potential (\ref{t77}) as a function of $\Sigma$ and $m_H$ has two GMPs that
are degenerate. One of them has the form $(m_H=-g/2,\Sigma=0)$, another looks like $(m_H=0,\Sigma=-g/2)$. The value 
of the function $V(m_H,\Sigma)$ at these points is the same and equal to $\frac{1}{48\pi}g^3<0$. The first GMP 
corresponds to the fact that only the Haldane mass term dynamically arises in the model (the other masses are equal to 
zero), and a phase occurs in which, e.g., parity $\cP_5$ is spontaneously broken down, although continuous 
$U(2N)$ chiral symmetry, including $U(1)_{\gamma^5}$ and $U(1)_{\gamma^3}$ chiral subgroups (\ref{n4}), remains intact.
In addition to these, there are other discrete symmetries of the model 
that are also spontaneously broken in this phase, or remain unbroken (for detailes, see in the Section \ref{II.1}). 
In the second GMP of the function (\ref{t77}) only the zero value of the Haldane mass term is fixed unambiguously, 
while the values of other masses $m_D,m_3,m_5$, which arises dynamically, are constrained by the condition 
\begin{eqnarray}
\Sigma^2\equiv m_D^2+m_5^2+m_3^2=g^2/4. \label{t78}
\end{eqnarray} 
Note that in each of the phases with $m_H=0$ and other masses satisfying the condition (\ref{t78}), at least one of 
the  chiral symmetries (\ref{n4}) is spontaneously broken. So, each phase from this variety is qualitatively different 
from the phase with $m_H=-g/2$ and $m_D=m_5=m_3=0$ in which chiral symmetries are not broken. However, it should 
be noted that there is much in common between all these 
dynamically arising at $g<0$ phases. Namely, the free energy density of their ground states is the same
and is equal to $\frac{1}{48\pi}g^3$, i.e. they are degenerate and can appear spontaneously 
in the massless (2+1)-D Thirring model (\ref{t1}) on the same footing. As a result, in the space,
filled with one of these degenerated phases, bubbles of the other phases can be created, i.e. one 
can observe in space the mixture (or coexistence) of these phases. The mass $M_F$ of the simplest quasiparticle 
excitations of their ground states, i.e. the pole of the fermion propagator (\ref{t14}), is also the same and  
$M_F=|g|/2$, etc. 

In this regard, we would like to note that the HF approach has recently been used to study the generalized (2+1)-D 
Thirring model containing both vector-vector and scalar-scalar interaction channels \cite{KGK2}. In this work, 
the possibility of dynamical generation of the Hermitian mass term of the form (\ref{t13}) was studied, but only with 
nonzero Dirac and Haldane mass terms. The results of this study, reduced to the case of the true Thirring model, 
turned out to be qualitatively the same as in the present investigation, which uses a more general mass term 
(\ref{t13}) with four different mass parameters. It means that in order to obtain the properties of the {\it Hermitian} 
ground state of the true (2+1)-D Thirring model, it is sufficient to use an ansatz (\ref{t13}) only with $m_D$ and 
$m_H$ masses. However (as it will be shown in the next section), in order to detect the phenomenon of spontaneous 
non-Hermiticity of the Thirring model, it is necessary to involve additional mass terms.

Finally, let us look at the phase structure of the model from the renormalization group point of view. For this we 
introduce dimensionless bare coupling constant $\lambda\equiv\lambda(\Lambda)=\Lambda G(\Lambda)$. This quantity is 
associated with the so-called Callan-Symanzik function $\beta (\lambda)=\Lambda\frac{\partial\lambda 
(\Lambda)}{\partial\Lambda}$. Using the relation (\ref{t73}), one can get
\begin{eqnarray}
\beta (\lambda)=\frac{\lambda}{\lambda_0}(\lambda_0-\lambda), \label{474}
\end{eqnarray}
where $\lambda_0=\frac{2N\pi^2}{3}$ is the nontrivial zero of the function $\beta (\lambda)$. The behavior of this 
function in the neighborhood of $\lambda_0$
indicates that $\lambda_0$ is an UV-stable point of the model. This means that in the continuous limit (that is, 
at $\Lambda\to\infty$) $\lambda (\Lambda)$ tends to the UV-stable point $\lambda_0$. (This feature of the 
dimensionless coupling constant $\lambda (\Lambda)$ can also be seen directly from Eq. (\ref{t75}).) Then, it is well
known that for values of $\lambda>\lambda_0$ a phase with broken symmetry is usually located,
and for $\lambda<\lambda_0$ a phase with unbroken symmetry occurs. These most general properties of the UV-stable 
point of any model are confirmed by the above calculations made in the framework of the massless (2+1)-D Thirring 
model. Indeed, it is easy to see from the relation (\ref{t75}) that
\begin{eqnarray}
\lambda-\lambda_0=-\frac{\pi^3Ng}{6\Lambda}+\cdots.\label{m760}
\end{eqnarray}
Therefore, if $g>0$, then, as it follows from Eq. (\ref{m760}), we have $\lambda <\lambda_0$. But when $g>0$, as is 
clear
from the previous discussion, the original symmetry of the model remains intact, i.e., the symmetrical phase of the 
model is located at $\lambda<\lambda_0$. But when $g<0$, any of the fermionic mass terms ($m_D, m_H,...$) can 
dynamically appear, and hence the spontaneous breaking of the original symmetry is realized. In this case, as can be 
seen from Eq. (\ref{m760}), we have $\lambda>\lambda_0$.

Two conclusions follow from this. (i) Within the HF approach to the (2+1)-D massless Thirring model built from 
reducible four-component spinors, fermion mass 
generation is possible for any finite value of $N$. (ii) It is obvious that $\lambda_0\to \infty$ when $N\to\infty$. 
Hence, in this case (at $N\to\infty$), for any finite values of $\lambda$ it is impossible to detect in the model 
(since $\lambda<\lambda_0=\infty$) the dynamical generation of different mass terms within 
the leading order of the $1/N$ approximation. The similar property has been predicted in some papers earlier, e.g., 
in Refs. \cite{Hong,Hyun}.

\section{Dynamical generation of the non-Hermitian mass term}\label{IV}

In the present section, we study in the HF approximation the possibility of the existence of such a solution 
$\overline S(p)$ of the HF equation (\ref{n043}), which would correspond to dynamical generation in the model of 
a non-Hermitian mass term ${\cal M}_{NH}$ of a rather general form,
\begin{eqnarray}
{\cal M}_{NH}=\overline\Psi_k (m_H\tau+\eta\cdot m_D+\vartheta\cdot im_5\gamma^5+\kappa\cdot im_3\gamma^3) \Psi_k,
\label{t0140}
\end{eqnarray}
where each of the multipliers $\eta, \vartheta, \kappa$ is either $1$ or $i$ and all mass parameters 
$m_H,m_D,m_5,m_3$ are real quantities. Note that the Haldane contribution to ${\cal M}_{NH}$, i.e. the term 
$m_H\overline\Psi_k \tau \Psi_k$, is a Hermitian \footnote{The non-Hermiticity of the Haldane mass term means that 
$m_H$ is not real. But in this case, as it is clear from Eq. (\ref{t16}), the spectrum of quasiparticles becomes 
complex, and the ground state of the system becomes unstable. So, throughout a paper we do not consider such a 
possibility. \label{f4}}, but the contributions from other mass terms, Dirac or chiral ones,
to the expression (\ref{t0140}) are not necessarily Hermitian. However, if all factors $\eta, \vartheta, \kappa$ in 
Eq. (\ref{t0140}) are equal to $1$, then ${\cal M}_{NH}$ as a whole becomes Hermitian (see in Eq. (\ref{t13})), 
and its generation within the framework of the HF approach was considered in the previous section \ref{III}.  
Therefore, here we assume that in Eq. (\ref{t0140}) at least one of the factors $\eta, \vartheta, \kappa$ is equal to $i$. In this case we 
must look for the solution $\overline S(p)$ of the gap HF equation 
(\ref{n043}) such that 
\begin{eqnarray}
\overline{S^{-1}}(p)=i\left (\hat p+m_H\tau+\eta\cdot m_D+\vartheta\cdot im_5\gamma^5+
\kappa\cdot im_3\gamma^3\right ). \label{t1400}
\end{eqnarray}
As for $\overline{S}(p)$ itself, in this case, one can use for it the expressions (\ref{t14})-(\ref{t16}) in which it 
is necessary to perform a replacements 
\begin{eqnarray}
m_D\to\eta m_D,~~m_3\to\kappa m_3,~~m_5\to\vartheta m_5,\label{t10}
\end{eqnarray}
and, as a consequence, to use there instead of $\Sigma$ the expression $\widetilde\Sigma$, i.e.
\begin{eqnarray}
\Sigma\to\widetilde\Sigma=\sqrt{\eta^2m_D^2+\vartheta^2 m_5^2+\kappa^2 m_3^2}.\label{t11}
\end{eqnarray}
Since we are looking for non-Hermitian phases with real spectrum of its quasiparticles (note, the energy spectrum of 
quasiparticles is defined by the singularity of the fermion propagator $\overline{S}(p)$), it must be supposed that 
$\eta, \vartheta, \kappa$ and $m_D,m_5,m_3$ are such that 
\begin{eqnarray}
\eta^2m_D^2+\vartheta^2 m_5^2+\kappa^2 m_3^2\ge 0.\label{t12}
\end{eqnarray}
Performing in Eq. (\ref{t18}) the same changes as in (\ref{t10}) and (\ref{t11}), it is easy to obtain the 
{\it unrenormalized} HF equations for 
$m_H,m_D,m_5$ and $m_3$ in the case of non-Hermitian mass term (\ref{t0140}). The UV-divergences of these equations 
can be removed by the same behavior of the coupling constant $G$ vs $\Lambda$ as in Eqs. (\ref{t73}) or (\ref{t75}). 
Hence, in the non-Hermitian case the {\it renormalized} HF equations for mass parameters $m_H,m_D,...$  look like Eq. 
(\ref{t74}), in which $\Sigma$ should be replaced by $\widetilde\Sigma$ according to Eq. (\ref{t11}). Of course, these 
equations are the stationary (or gap) equations of the HF effective potential $V_{NH}(m_H,m_D,m_5,m_3)$ in the 
non-Hermitian case. It can be obtained on the basis of a general expression (\ref{t76}) in the same way as the HF 
effective potential (\ref{t77}) in the Hermitian case (see in Appendix \ref{ApB}) and has the form
\begin{eqnarray}
V_{NH}(m_H,m_D,m_5,m_3)\equiv V_{NH}(m_H,\widetilde\Sigma) =\frac{1}{12\pi}\Big (3 g\widetilde\Sigma^2+3 g m_H^2 
+2|\widetilde\Sigma+m_H|^3+2 |\widetilde\Sigma-m_H|^3\Big ),\label{t9}
\end{eqnarray}
where $\widetilde\Sigma=\sqrt{\eta^2m_D^2+\vartheta^2 m_5^2+\kappa^2 m_3^2}\ge 0$ and $g$ is defined in Eq. (\ref{t73}).
The GMP of this function over the variables $m_H,m_D,m_5$ and $m_3$ defines the massive components of the non-Hermitian 
mass term (\ref{t0140}) that can appear dynamically in the model and, as a result, the properties of the corresponding 
non-Hermitian ground state of the model. 

\subsection{The case $g>0$: gapless spontaneous non-Hermiticity of the (2+1)-dimensional Thirring model}

In this case the effective potential (\ref{t9}), considered as a function  of $m_H$ and 
$\widetilde\Sigma$, has a trivial GMP of the form $(m_H=0,\widetilde\Sigma=0)$. However, we must look at 
$V_{NH}(m_H,\widetilde\Sigma)$, where $\widetilde\Sigma=\sqrt{\eta^2m_D^2+\vartheta^2 m_5^2+\kappa^2 m_3^2}$, 
as a composite function of $m_D,m_3,m_5$ variables, constrained by the relation 
$\eta^2m_D^2+\vartheta^2 m_5^2+\kappa^2 m_3^2\ge 0$. 
It is evident that in the Hermitian case, i.e. when $\eta=\vartheta=\kappa=1$,
there is no dynamical mass generation in the model, since the GMP of its effective potential (\ref{t9}) 
looks like $m_H=m_D=m_5=m_3=0$, where we have $V_{NH}(0,0,0,0)=0$. And there are no other points $(m_H,m_D,m_5,m_3)$,
different from this trivial point, at which this relation would hold. 

In contrast, if one or two of the factors $\eta,\vartheta,\kappa$ 
are equal to $i$, then in the four-dimensional $(m_H,m_D,m_5,m_3)$ space there are nontrivial two-dimensional
manifolds, on which the effective potential $V_{NH}$ (\ref{t9}) vanishes, i.e. reaches its lowest value. Namely,
in this case for each fixed set $\eta,...$ the corresponding two-dimensional manifold is defined by two equations, 
$m_H=0$ and $\widetilde\Sigma=0$. Each nontrivial point of this manifold is also a GMP of the effective potential 
(\ref{t9}) which defines the dynamically generated non-Hermitian mass term of the model. 

For example, if $\eta=\vartheta=1,\kappa=i$ and $g>0$, then a non-Hermitian mass 
term $\overline\Psi_k(m_D+im_5\gamma^5-m_3\gamma^3) \Psi_k$ with $m_H=0$ can be generated, in which nonzero 
masses satisfy the relation 
$\widetilde\Sigma=0$, i.e. $m_D^2 + m_5^2=m_3^2$. This mass term corresponds to the GMP of the effective potential 
(\ref{t9}) where 
it is equal to zero. It is clear from Table \ref{Tab1} that in the phase with such a mass term the discrete 
$\cP_5\cT_5$ and $\cP_3\cT_3$ symmetries remain intact, but other $\cP_k\cT_l$ symmetries are spontaneously broken.
Moreover, it follows from Eq. (\ref{t1400}) that the corresponding fermion propagator $\overline {S}(p)$ looks like
\begin{eqnarray}
\overline {S}(p)=-i(\hat p+m_D+i\gamma^5 m_5-\gamma^3 m_3)/p^2. \label{472}
\end{eqnarray}
The relation (\ref{472}) means that quasiparticle excitations of this phase are massless and, hence, their energy 
spectrum has zero excitation energy (or zero gap). In the literature, a phase in which quasiparticles are gapless 
is usually called gapless (for example, there are gapless color superconductivity \cite{Huang} and gapless charged 
pion condensation \cite{EK}, etc.). By analogy, this non-Hermitian phase of the (2+1)-D Thirring model can also be
called gapless. 

In addition to the above phase, at $g>0$ five other non-Hermitian gapless phases (corresponding to other 
$\eta,\vartheta,\kappa$ sets) may appear spontaneously in the model. They are characterized by dynamical 
appearance of one of the following non-Hermitian mass terms (with $m_H=0$),
\begin{eqnarray}
&&({\rm i})~~~\overline\Psi_k(m_D-m_5\gamma^5+im_3\gamma^3)\Psi_k~~~{\rm where}~~~m_D^2+ m_3^2= m_5^2,\nonumber\\
&&({\rm ii})~~\overline\Psi_k(im_D+im_5\gamma^5+im_3\gamma^3) \Psi_k~~~{\rm where}~~~m_5^2+ m_3^2= m_D^2,\nonumber\\
&&({\rm iii})~~\overline\Psi_k(im_D-m_5\gamma^5+im_3\gamma^3)\Psi_k~~~{\rm where}~~~m_D^2+ m_5^2= m_3^2,\nonumber\\
&&({\rm iv})~~\overline\Psi_k(im_D+im_5\gamma^5-m_3\gamma^3) \Psi_k~~~{\rm where}~~~m_D^2+ m_3^2= m_5^2,\nonumber\\
&&({\rm v})~~~\overline\Psi_k(m_D-m_5\gamma^5-m_3\gamma^3) \Psi_k~~~{\rm where}~~~m_5^2+ m_3^2= m_D^2. \label{t20}
\end{eqnarray}
With respect to the transformations $\cP_5\cT_3$ and $\cP_3\cT_5$ the mass term (i) is even, while (iv) 
is odd. The mass term (iii)
is $\cP_5\cT_5$ and $\cP_3\cT_3$ odd, whereas the mass terms (ii) and (v) are neither symmetric, nor antisymmetric
under any of the $\cP_k\cT_l$ discrete transformations (see in Table \ref{Tab1}). Note that all these six gapless 
non-Hermitian phases are degenerated, since the energy density of the ground state in each of these phases is the same
and equals zero. Moreover, at $g>0$ the Hermitian symmetric phase, in which all masses $m_D,m_H,...$ 
are equal to zero (see in the previous section \ref{III}), 
also has a zero free energy density and, as a result, can appear in the system on an equal footing with all 
these gapless non-Hermitian phases.

\subsection{The case $g<0$}

Just as discussed in the section \ref{III}, in this case the HF effective potential (\ref{t9})
as a function of $m_H$ and $\widetilde\Sigma$ has two degenerated GMPs, $(m_H=-g/2,\widetilde\Sigma=0)$ and 
$(m_H=0,\widetilde\Sigma=-g/2)$.
Excluding again from a consideration the possibility of dynamical generation of the Hermitian mass term, 
i.e. the case when $\eta=\vartheta=\kappa=1$, we see that the first GMP corresponds to the possibility 
of spontaneous generation in the Thirring model of six different non-Hermitian phases corresponding to the following 
dynamically generated non-Hermitian mass terms
\begin{eqnarray}
&&({\rm i})~~~~\overline\Psi_k(m_H\tau+m_D+im_5\gamma^5-m_3\gamma^3)\Psi_k~~~{\rm with}~~~m_D^2+ m_5^2= m_3^2,\nonumber\\
&&({\rm ii})~~~\overline\Psi_k(m_H\tau+m_D-m_5\gamma^5+im_3\gamma^3)\Psi_k~~~{\rm with}~~~m_D^2+ m_3^2= m_5^2,\nonumber\\
&&({\rm iii})~~\overline\Psi_k(m_H\tau+im_D-m_5\gamma^5+im_3\gamma^3)\Psi_k~~~{\rm with}~~~m_D^2+ m_5^2= m_3^2,\nonumber\\
&&({\rm iv})~~\overline\Psi_k(m_H\tau+im_D+im_5\gamma^5-m_3\gamma^3) \Psi_k~~~{\rm with}~~~m_D^2+ m_3^2= m_5^2,\nonumber\\
&&({\rm v})~~~\overline\Psi_k(m_H\tau+im_D+im_5\gamma^5+im_3\gamma^3) \Psi_k~~~{\rm with}~~~m_5^2+ m_3^2= m_D^2,\nonumber\\
&&({\rm vi})~~\overline\Psi_k(m_H\tau+m_D-m_5\gamma^5-m_3\gamma^3) \Psi_k~~~{\rm with}~~~m_5^2+ m_3^2= m_D^2, \label{t21}
\end{eqnarray}
where $m_H=-g/2$ and $\widetilde\Sigma =0$. Each of the non-Hermitian mass terms (i)-(vi) of Eq. (\ref{t21}) is assigned to a well-defined set 
of parameters $\eta,\vartheta,\kappa$. For example, the mass term (i) corresponds to $\eta=\vartheta=1,\kappa=i$, 
the mass term (ii) corresponds to $\eta=\kappa=1, \vartheta=i$, etc.
It follows from Table \ref{Tab1} that the mass term (i) of Eq. (\ref{t21}) and the corresponding non-Hermitian phase are 
$\cP_5\cT_5$ invariant. Moreover, the phase with mass term (ii) is $\cP_5\cT_3$ symmetric. It is interesting to note
that the mass terms (iii) and (iv) are, respectively, $\cP_3\cT_3$ and $\cP_3\cT_5$ odd, but the rest mass terms,
(v) and (vi), have no any $\cP_k\cT_l$ parity. 

The second GMP of the HF effective potential $V_{NH}(m_H,\widetilde\Sigma)$ (\ref{t9}), i.e. the point 
$(m_H=0,\widetilde\Sigma=-g/2)$, corresponds to other six non-Hermitian phases of the model with the following 
dynamically generated non-Hermitian mass terms
\begin{eqnarray}
&&({\rm i})~~~~\overline\Psi_k(m_D+im_5\gamma^5-m_3\gamma^3)\Psi_k~~~{\rm with}~~~m_D^2+ m_5^2-m_3^2=g^2/4,\nonumber\\
&&({\rm ii})~~~\overline\Psi_k(m_D-m_5\gamma^5+im_3\gamma^3)\Psi_k~~~{\rm with}~~~m_D^2+ m_3^2-m_5^2=g^2/4,\nonumber\\
&&({\rm iii})~~\overline\Psi_k(im_D+im_5\gamma^5+im_3\gamma^3) \Psi_k~~~{\rm with}~~~m_5^2+ m_3^2-m_D^2=g^2/4,\nonumber\\
&&({\rm iv})~~\overline\Psi_k(im_D-m_5\gamma^5+im_3\gamma^3)\Psi_k~~~{\rm with}~~~m_3^2-m_D^2-m_5^2=g^2/4,\nonumber\\
&&({\rm v})~~~\overline\Psi_k(im_D+im_5\gamma^5-m_3\gamma^3) \Psi_k~~~{\rm with}~~~m_5^2-m_D^2-m_3^2=g^2/4,\nonumber\\
&&({\rm vi})~~\overline\Psi_k(m_D-m_5\gamma^5-m_3\gamma^3) \Psi_k~~~{\rm with}~~~m_D^2-m_5^2-m_3^2=g^2/4.\label{t22}
\end{eqnarray}
In Eq. (\ref{t22}), the non-Hermitian phase corresponding to the mass term (i) is symmetric under $\cP_5\cT_5$ and $\cP_3\cT_3$, whereas the 
phase with dynamically arising mass term (ii) is invariant under $\cP_5\cT_3$ and $\cP_3\cT_5$ transformations. In 
contrast, in any of the non-Hermitian phases with one of the (iii)-(vi) mass terms, all discrete symmetries discussed 
in Section \ref{II.1} are spontaneously broken. However, it is worth clarifying that the mass term (iv) 
is $\cP_5\cT_5$ and $\cP_3\cT_3$ odd, while the mass term (v) is $\cP_5\cT_3$ and $\cP_3\cT_5$ odd.  

It is easy to see that at $g<0$ the fermion propagator $\overline S(p)$ of Eq. (\ref{t1400}) in each of the non-Hermitian
phases, corresponding to (non-Hermitian) mass terms of Eqs. (\ref{t21}) and (\ref{t22}), describes the quasiparticles
with the same real mass $M_F=|g|/2$. (Note that it is also the same as the mass of quasiparticle excitations of 
all Hermitian phases which can dynamically arise at $g<0$ (see in Sec. \ref{III}).) 
Moreover, all non-Hermitian and Hermitian phases, which appear dynamically in the (2+1)-D Thirring model at 
$g<0$, are degenerated since the density of their ground state energies is the same and equals $g^3/(48\pi)$.

Recall that in the recent papers \cite{Chernodub2,KKZ} it has been proven that spontaneous (and spatially
homogeneous) non-Hermiticity 
found in some {\it massless} models with four-fermion interactions disappears if a nonzero bare (Hermitian) 
fermion mass is introduced into the model. In contrast, in the next section we show that the effect of spontaneous 
non-Hermiticity, observed in massless (2+1)-D Thirring model, can also occur in the {\it massive} variant of the model.

\section{Spontaneous non-Hermiticity in the massive Thirring model}
\subsection{The case of nonzero bare Haldane mass}

First of all, we investigate the question of the possibility of spontaneous non-Hermiticity in the framework 
of the massive (2+1)-D Thirring model with a nonzero bare Haldane mass $m_{0H}$. In this case the Lagrangian of 
the model looks like 
\begin{eqnarray}
 L=\overline \Psi_k\big [\gamma^\nu i\partial_\nu + \tau m_{0H}\big ]\Psi_k&-& \frac {G}{2N}\left
(\overline \Psi_k\gamma^\mu\Psi_k\right )\left (\overline \Psi_l\gamma_\mu\Psi_l\right ).\label{m01}
\end{eqnarray}
This Lagrangian is invariant under continuous chiral transformations (\ref{n4}), discrete space parity $\cP_3$ and two $\cP\cT$ transformations, 
$\cP_5\cT_5$ and $\cP_5\cT_3$ (see in Table \ref{Tab1}).
Other discrete symmetries of the massless Thirring model (\ref{t1}), which have been considered in Sec. \ref{II.1},
are violated explicitly by the Haldane mass term. However, the continuous $U(2N)$ 
symmetry inherent in the model (\ref{t1}) is also characteristic of Lagrangian (\ref{m01}). To begin with, we are going 
to study the possibility of dynamical appearance of different Hermitian 
mass terms, $m_D$, $m_3$ and $m_5$ (in addition to Haldane mass $m_H$), in the framework of the HF approach to this
massive model.

\subsubsection{Dynamical generation of Hermitian mass terms}

The consideration is again performed on the basis of the CJT effective action (\ref{n420}) and its
stationary HF equation (\ref{n0420}), in which this time $D(x,y)=[\gamma^\nu i\partial_\nu+
 \tau m_{0H}]\delta^3(x-y)$, i.e. $\overline D(p)=\hat p+ \tau m_{0H}$.
Fourier transformation of the gap HF equation (\ref{n0420}) reads as 
\begin{eqnarray}
-i\overline{(S^{-1})^\beta_\alpha}(p) &=&(\hat p)^\beta_\alpha+m_{0H}(\tau)^\beta_\alpha
-G\big (\gamma^\rho\big )^\beta_\alpha\int\frac{d^3q}{(2\pi)^3}~{\rm tr}
\big [\gamma_\rho\overline{S}(q)\big ]
+\frac{G}N \int\frac{d^3q}{(2\pi)^3}~\big [\gamma^\rho\overline{S}(q)\gamma_\rho\big ]^\beta_\alpha.
 \label{n0431}
\end{eqnarray}
We are looking for the solution $\overline{S}(p)$ of the gap HF equation (\ref{n0431}) in the form
presented by Eqs. (\ref{t14})-(\ref{t16}), i.e. 
\begin{eqnarray}
\overline{S^{-1}}(p)=i(\hat p+m_D+m_H\tau+i\gamma^5 m_5+i\gamma^3 m_3), \label{m03}
\end{eqnarray}
where $m_D$, $m_H$, $m_3$ and $m_5$ are real quantities (it corresponds to dynamical appearance of the Hermitian mass
term (\ref{t13}) in the model (\ref{m01})). Using this ansatz in Eq. (\ref{n0431}), one can obtain 
for $m_D$, $m_H$, $m_3$ and $m_5$ the {\it unrenormalized} system of gap equations in which all equations look the 
same as in the system (\ref{t18}), except for the first one, which has the form 
\begin{eqnarray}
m_H&=&m_{0H}+\frac{3iG}{2N}\int\frac{d^3p}{(2\pi)^3}\left\{
\frac{\Sigma+m_H}{p^2-(\Sigma+m_H)^2}-\frac{\Sigma-m_H}{p^2-(\Sigma-m_H)^2}\right\}.\label{t180}
\end{eqnarray}
To renormalize the obtained system of gap equations, we require (i) that the bare coupling constant $G$ has the same dependence 
(\ref{t75}) on the cutoff parameter $\Lambda$ as in the massless case. (ii) Moreover, we need at $\Lambda\to\infty$
the fulfillment of the following relation 
\begin{eqnarray}
\frac{m_{0H}}{G}=\pm\frac{3\mu^2}{8N\pi}+\mu^2{\cal O}\left (\frac \mu\Lambda\right ), \label{m041}
\end{eqnarray}
where, in addition to $g$, $\mu$ is another finite and renormalization group invariant free model parameter. 
Like $g$, it has the dimension of [mass]. For definiteness, in the following consideration we select the 'plus' sign
in Eq. (\ref{m041}). In this case, just under the above (i) and (ii) constraints on $G$ and $m_{0H}$, the {\it renormalized} 
HF system of gap equations for $m_D$, $m_H$, $m_3$ and $m_5$ can be obtained at $\Lambda\to\infty$,
\begin{eqnarray}
&&gm_H+ (\Sigma+m_H)|\Sigma+m_H|-(\Sigma-m_H)|\Sigma-m_H|=\mu^2,\nonumber\\
&&m_D\Big\{g+ \frac 1\Sigma\Big[(\Sigma+m_H)|\Sigma+m_H|+(\Sigma-m_H)|\Sigma-m_H|\Big]\Big\}=0,\nonumber\\
&&m_3\Big\{g+ \frac 1\Sigma\Big[(\Sigma+m_H)|\Sigma+m_H|+(\Sigma-m_H)|\Sigma-m_H|\Big]\Big\}=0,\nonumber\\
&&m_5\Big\{g+ \frac 1\Sigma\Big[(\Sigma+m_H)|\Sigma+m_H|+(\Sigma-m_H)|\Sigma-m_H|\Big]\Big\}=0.\label{t741}
\end{eqnarray}
In general, 
the stationary HF system of 
equations (\ref{t741}) has several solutions. And the most preferred of them is the one that minimizes the value of 
the CJT effective potential $V(S)$ defined in Eq. (\ref{t76}). In the HF approach to the massive Thirring model 
(\ref{m01}), it looks like
\begin{eqnarray}
V(S)\equiv V_H^{mas}(m_H,\Sigma)=\frac{1}{12\pi}\Big (- 6\mu^2m_H+3 g\Sigma^2+3 g m_H^2 +2|\Sigma+m_H|^3+
2 |\Sigma-m_H|^3\Big ),\label{B20}
\end{eqnarray}
where $\Sigma=\sqrt{m_D^2+m_3^2+m_5^2}$ and subscript $H$ means that effective potential $V_H^{mas}$ corresponds to 
a model (\ref{m01}) with bare Haldane mass. (The expression (\ref{B20}) is obtained in Appendix \ref{ApB2}.)

Suppose that $g>0$. In this case both terms in curly braces of the system (\ref{t741}) are positive, 
so its solution is such that $m_D=m_3=m_5=0$, i.e. with $\Sigma=0$. Then a remaining mass, $m_H$, obeys the equation
\begin{eqnarray}
&&gm_H+ 2m_H|m_H|=\mu^2.\label{t751}
\end{eqnarray}
Hence, at $g>0$ the HF system of equations (\ref{t741}) has a single solution with $m_D=m_3=m_5=0$. Moreover,
its Haldane mass component looks like
\begin{eqnarray}
&&m_H=(m_{H})_0\equiv (\sqrt{g^2+8\mu^2}-g)/4.\label{t752}
\end{eqnarray}
But in the case $g<0$, depending on the relationship between $g$ and $\mu$, the stationary HF system of 
equations (\ref{t741}) has several solutions. And the most preferred of them is the one that minimizes the value of 
the CJT effective potential (\ref{B20}). Investigating this function with the help of any program of analytical 
calculations, it can be shown that in the GMP of the effective potential (\ref{B20}) we have again $m_D=m_3=m_5=0$ and 
$m_H=(m_{H})_0$ (the last quantity is also defined by Eq. (\ref{t752})). 

As a result, we see that no other Hermitian mass terms are dynamically generated in the model (\ref{m01}) 
in addition to Haldane mass. The Hermitian ground state of this massive Thirring model has the 
same symmetry as the initial bare Lagrangian (\ref{m01}), and the mass of its quasiparticle excitations, i.e. 
the pole of the propagator $\overline{S}(p)$ (\ref{m03}), equals $(m_{H})_0$ (\ref{t752}). \label{VA1}

\subsubsection{Spontaneous non-Hermiticity}

Let us now consider the possibility of a situation when a non-Hermitian ground state can be realized in a perfectly 
Hermitian massive model (\ref{m01}). In this case, the dynamics of quasiparticle excitations of such a ground state 
is effectively described by non-Hermitian Lagrangian. Within the framework of the HF approach to the model (\ref{m01}), 
it is possible to find a dynamically generated non-Hermitian mass term ${\cal M}_{NH}$ of this effective Lagrangian. 
It looks like an expression (\ref{t0140}), in which at least one of the factors $\eta,\vartheta,\kappa$ is necessarily 
equal to $i$, and the others are units. To find ${\cal M}_{NH}$, we should look for a solution of the HF equation 
(\ref{n0431}) in the form (\ref{t1400}). Further, assuming the same asymptotic expansions (\ref{t75}) and (\ref{m042}),
respectively, for the bare model parameters $G$ and $m_{0H}$, we can obtain a renormalized system of HF equations 
for the parameters $m_H,m_D,m_3$ and $m_5$ appearing in the non-Hermitian mass term (\ref{t0140}). 
It has the same form as the system of HF equations (\ref{t741}), in which it is necessary to substitute 
$\widetilde\Sigma$ instead of $\Sigma$ (recall, $\widetilde\Sigma=\sqrt{\eta^2m_D^2+\vartheta^2
m_5^2+\kappa^2 m_3^2}$). Similarly, one can obtain the effective potential in the HF approximation which describes 
the non-Hermitian ground state of the massive (2+1)-D Thirring model (\ref{m01}). It looks like 
\begin{eqnarray}
V_{NHH}^{mas}(m_H,m_D,m_3,m_5)=\frac{1}{12\pi}\Big (- 6\mu^2m_H+3 g\widetilde\Sigma^2+3 g m_H^2 +2|\widetilde\Sigma+m_H|^3+
2 |\widetilde\Sigma-m_H|^3\Big ).\label{B201}
\end{eqnarray}
And just its GMPs provide information both about the symmetry properties of the non-Hermitian ground state of the 
model and about the mass spectrum of its quasiparticles. 

Effective potential (\ref{B201}) as a function of $m_H$ and $\widetilde\Sigma$ has a GMP of the form $(m_H=(m_{H})_0,
\widetilde\Sigma=0)$ (see the discussion on effective potential properties (\ref{B20}) in the section \ref{VA1}).
As a result, we see that, in addition to the trivial case with $\eta=\vartheta=\kappa=1$ (it corresponds to a dynamical 
generation of the Hermitian mass term considered in the previous section \ref{VA1}), there are six different sets 
$\eta,\vartheta,\kappa$, and for each one the effective potential (\ref{B201}), as a function of 
$m_H,m_D,m_3$ and $m_5$, reaches its smallest value (the same one) at each point of a two-dimensional manifold 
of the form $m_H=(m_{H})_0,\widetilde\Sigma=\sqrt{\eta^2m_D^2+\vartheta^2m_5^2+\kappa^2 m_3^2}=0$. Each point of 
this manifold corresponds to the non-Hermitian ground state of the system described effectively by Lagrangian with 
non-Hermitian mass term of the form (\ref{t0140}). At arbitrary sign of $g$, all possible non-Hermitian mass terms 
dynamically generated in this case in the model are listed in Eq. (\ref{t21}) in which $m_H=(m_{H})_0$ of Eq. 
(\ref{t752}). They appear
in the model spontaneously, and the corresponding non-Hermitian phases are degenerated since their ground state 
energy is the same. In addition, all these non-Hermitian phases are degenerated with Hermitian phase corresponding to 
the case $\eta=\vartheta=\kappa=1$, where $m_D=m_3=m_5=0$ and $m_H=(m_{H})_0$ (\ref{t752}). \label{VA2}

\subsection{The case of nonzero bare Dirac mass}
Let us explore another limiting variant of the massive (2+1)-D Thirring model, when Lagrangian (\ref{t1}) is 
extended by the Hermitian bare Dirac mass term of the form $m_{0D}\overline \Psi_k \Psi_k$.
In this case, massive Thirring model is invariant under all discrete transformations considered in Sec. \ref{II.1}. 
However, the continuous chiral symmetries (\ref{n4}) of the massless model are violated explicitly by the bare Dirac 
mass term. As a result, in this case the $U(2N)$ symmetry of the massless model (\ref{t1}) is reduced to $U(N)$. Here,
we are going to study in the framework of the HF approach to this massive model the possibility of dynamical 
generation of both Hermitian and non-Hermitian mass terms in it.

Again, our consideration is performed on the basis of the CJT effective action (\ref{n420}) and its
stationary HF equation (\ref{n0420}), in which this time $D(x,y)=[\gamma^\nu i\partial_\nu+
 m_{0D}]\delta^3(x-y)$, i.e. $\overline D(p)=\hat p+  m_{0D}$.
Fourier transformation of this gap HF equation (\ref{n0420}) reads as 
\begin{eqnarray}
-i\overline{(S^{-1})^\beta_\alpha}(p) &=&(\hat p)^\beta_\alpha+m_{0D}\delta^\beta_\alpha
-G\big (\gamma^\rho\big )^\beta_\alpha\int\frac{d^3q}{(2\pi)^3}~{\rm tr}
\big [\gamma_\rho\overline{S}(q)\big ]
+\frac{G}N \int\frac{d^3q}{(2\pi)^3}~\big [\gamma^\rho\overline{S}(q)\gamma_\rho\big ]^\beta_\alpha.
 \label{n0432}
\end{eqnarray}

$\bullet$ First, we consider the possibility that the solution $\overline{S}(p)$ of the gap HF equation (\ref{n0432}) 
corresponds to a Hermitian ground state of the model. In this case, it has the form presented by Eqs. 
(\ref{t14})-(\ref{t16}) with real $m_D$, $m_H$, $m_3$ and $m_5$ masses and corresponds to Hermitian mass term 
(\ref{t13}) which can be generated dynamically. Using this ansatz in Eq. (\ref{n0432}), one can obtain 
for $m_D$, $m_H$, $m_3$ and $m_5$ the {\it unrenormalized} system of gap equations in which all equations look the 
same as in the system (\ref{t18}), except this time for the second one, which has the form 
\begin{eqnarray}
m_D&=&m_{0D}+\frac{3iG}{2N}\frac{m_D}{\Sigma}\int\frac{d^3p}{(2\pi)^3}\left\{
\frac{\Sigma+m_H}{p^2-(\Sigma+m_H)^2}+\frac{\Sigma-m_H}{p^2-(\Sigma-m_H)^2}\right\}.\label{t181}
\end{eqnarray}
To renormalize this system of equations, we again require (i) that the bare coupling constant $G$ has the same 
dependence 
(\ref{t75}) on the cutoff parameter $\Lambda$ as in the massless case. (ii) Moreover, we need at $\Lambda\to\infty$
the fulfillment of the following relation 
\begin{eqnarray}
\frac{m_{0D}}{G}=\pm\frac{3\mu^2}{8N\pi}+\mu^2{\cal O}\left (\frac \mu\Lambda\right ), \label{m042}
\end{eqnarray}
where, in addition to $g$, $\mu$ is another finite and renormalization group invariant free model parameter. 
Like $g$, it has the dimension of [mass]. For definiteness, we will further consider the situation when the expression 
(\ref{m042}) contains a 'plus' sign. Then, taking into account (i) and (ii) constraints on bare parameters, the 
{\it renormalized} HF system of gap equations for $m_D$, $m_H$, $m_3$ and $m_5$ can be obtained at $\Lambda\to\infty$
in the Hermitian case,
\begin{eqnarray}
&&gm_H+ (\Sigma+m_H)|\Sigma+m_H|-(\Sigma-m_H)|\Sigma-m_H|=0,\nonumber\\
&&m_D\Big\{g+ \frac 1\Sigma\Big[(\Sigma+m_H)|\Sigma+m_H|+(\Sigma-m_H)|\Sigma-m_H|\Big]\Big\}=\mu^2,\nonumber\\
&&m_3\Big\{g+ \frac 1\Sigma\Big[(\Sigma+m_H)|\Sigma+m_H|+(\Sigma-m_H)|\Sigma-m_H|\Big]\Big\}=0,\nonumber\\
&&m_5\Big\{g+ \frac 1\Sigma\Big[(\Sigma+m_H)|\Sigma+m_H|+(\Sigma-m_H)|\Sigma-m_H|\Big]\Big\}=0
\label{t742}
\end{eqnarray}
where $\Sigma=\sqrt{m_D^2+m_3^2+m_5^2}$. And we should find such a solution of the gap system (\ref{t742}) that 
minimizes the CJT effective potential $V(S)$ (\ref{t76}) calculated in the HF approximation,
\begin{eqnarray}
V(S)\equiv V_D^{mas}(m_D,m_H,\Sigma)=\frac{1}{12\pi}\Big (- 6\mu^2m_D+3 g\Sigma^2+3 g m_H^2 +2|\Sigma+m_H|^3+
2 |\Sigma-m_H|^3\Big ).\label{B202}
\end{eqnarray}
(The expression (\ref{B202}) is obtained in Appendix \ref{ApB2}.)

Since in the present consideration we deal with nonzero values of the mass parameter $\mu$, it is clear from the 
second of equations (\ref{t742}) that for both signs of $g$ the expression in curly braces must be nonzero. So it 
follows from the last two equations of this system that $m_3=m_5=0$, i.e. $\Sigma=|m_D|$. And the GMP 
of the function (\ref{B202}) must satisfy this condition. With this in mind, it is much more 
convenient to study the effective potential (\ref{B202}) with the help of any program of analytical calculations. 
As a result, one can see that at the GMP of the function $V_D^{mas}$ we have $m_H=m_3=m_5=0$ and 
$m_D=(m_{D})_0\equiv (\sqrt{g^2+8\mu^2}-g)/4$. It means that the Hermitian ground state of this kind of the 
massive Thirring model has the same symmetry as the initial Lagrangian with nonzero bare Dirac mass. And the mass 
of its quasiparticle excitations, i.e. the pole of the propagator $\overline{S}(p)$, equals $(m_{D})_0$.

$\bullet$$\bullet$ Now, let us study the possibility that the solution $\overline{S}(p)$ of the gap HF equation 
(\ref{n0432}) corresponds to a non-Hermitian ground state spontaneously arising in the massive Thirring model 
with nonzero bare Dirac mass term. To this end, we are looking for the solution $\overline{S}(p)$ in such a form that
\begin{eqnarray}
\overline{S^{-1}}(p)=i\left (\hat p+m_H\tau+m_D+\vartheta\cdot im_5\gamma^5+
\kappa\cdot im_3\gamma^3\right ), \label{t15}
\end{eqnarray}
where each of the values $\vartheta, \kappa$ is either $1$ or $i$ and all mass parameters 
$m_H,m_D,m_5,m_3$ are real quantities. In this case the solution $\overline{S}(p)$ of the HF equation (\ref{n0432}) 
corresponds to a dynamical 
generation of the following mass term 
\begin{eqnarray}
{\cal M}_{NH}=\overline\Psi_k (m_H\tau+m_D+\vartheta\cdot im_5\gamma^5+\kappa\cdot im_3\gamma^3) \Psi_k,\label{t015}
\end{eqnarray}
which becomes non-Hermitian (at $m_{3,5}\ne 0$) if at least one of the factors $\vartheta, \kappa$ is equal 
to $i$. (Therefore, we assume that the Dirac and Haldane masses remain Hermitian, and the non-Hermiticity 
of the ground state arises spontaneously from the non-Hermitian chiral $m_{3,5}$-mass terms. A more detailed 
discussion why the Haldane and Dirac mass terms in Eq. (\ref{t15}) are selected to be Hermitian is presented at 
the end of the section.) 
Substituting expression (\ref{t15}) into HF equation (\ref{n0432}) and postulating there the same behaviors 
of the bare coupling constant $G$ and the bare Dirac mass $m_{0D}$, which are given in formulas (\ref{t75}) and 
(\ref{m042}), respectively, one can obtain for $m_D, m_H, m_3$ and $m_5$ a renormalized HF system of equations 
of the form (\ref{t742}), in which the following substitutions must be made,
$m_3\to\kappa m_3,~~m_5\to\vartheta m_5$ and $\Sigma\to\widetilde\Sigma=\sqrt{m_D^2+\vartheta^2 m_5^2+\kappa^2 m_3^2}$.
Since we are looking for non-Hermitian phases with real spectrum of its quasiparticles, 
it must be supposed that 
$\vartheta, \kappa$ and $m_D,m_5,m_3$ are such that $m_D^2+\vartheta^2 m_5^2+\kappa^2 m_3^2\ge 0$. Moreover, performing
the same replacements in Eq. (\ref{B202}), we obtain HF effective potential $V(S)$ of the massive Thirring model with nonzero bare Dirac mass $m_{0D}$ when fermion 
propagator $\overline{S}(p)$ looks like in Eq. (\ref{t15}), i.e. 
\begin{eqnarray}
V(S)\equiv V_{NHD}^{mas}(m_D,m_H,\widetilde\Sigma)=\frac{1}{12\pi}\Big (- 6\mu^2m_D+3 g\widetilde\Sigma^2+3 g m_H^2
+2|\widetilde\Sigma+m_H|^3+
2 |\widetilde\Sigma-m_H|^3\Big ).\label{B204}
\end{eqnarray}
Now we are ready to conclude that non-Hermitian mass term of the form (\ref{t015}) cannot be generated dynamically in
the massive Thirring model under consideration. The first reason for this is the fact that for any allowable values of 
$\vartheta, \kappa$ (recall that $\vartheta=1$ or $i$ and $\kappa=1$ or $i$), the corresponding system of HF equations
has only solutions with $m_3=m_5=0$ (see also the discussion after Eq. (\ref{B202}) in the Hermitian case), and, as a 
consequence, the non-Hermiticity in the mass term (\ref{t015}) disappears. 
Secondly, it is also interesting to note that in all cases when we try to find a non-Hermitian solution of the 
form (\ref{t15}) of the HF equation (\ref{n0432}), the corresponding effective potential (\ref{B204}) becomes 
unbounded from below as soon as $\vartheta\ne 1$ and/or $\kappa\ne 1$. To 
confirm this fact, consider for simplicity the case 
$\vartheta, \kappa=i$. As a result, the quantity $\widetilde\Sigma$ in Eq. (\ref{B204}) takes the form
$\widetilde\Sigma=\sqrt{m_D^2-m_3^2-m_5^2}$.
Then, supposing that in Eq. (\ref{B204}) $m_H=const$, $\widetilde\Sigma=0$, i.e. $m_D^2=m_3^2+m_5^2$, we see that
in the limit $m_D\to\infty$ the effective potential
$V_{NHD}^{mas}$ tends to $-\infty$, i.e. it is an unbounded from below function. Hence the non-Hermitian 
{\it stable} ground state which is characterized by non-Hermitian mass term (\ref{t015}) 
cannot be generated spontaneously in the (2+1)-D massive Thirring model with nonzero bare Dirac mass $m_{0D}$.

In this subsection, we have excluded from consideration the possibility of the dynamical appearance of the 
non-Hermitian Dirac and Haldane mass terms in the model (see in Eq. (\ref{t015})). The reason for Dirac mass $m_D$ is 
that, as it is easy to see from Eq. (\ref{B204}), the effective potential becomes a complex-valued quantity when 
$m_D$ is not real, and hence the ground state of the system is unstable. As regards the exclusion from consideration 
of the non-Hermitian Haldane mass term, 
in this case the spectrum of quasiparticles, i.e. the pole of 
the fermionic propagator $\overline{S}(p)$, becomes explicitly complex-valued, and the theory becomes unstable 
(see also the remark made in footnote \ref{f4} at the beginning of Section IV).\vspace{0.3cm} 

The main result of this section is the following. Up to now, the phenomenon of spontaneous non-Hermiticity has been 
established only in the frameworks of some {\it massless} quantum field theory models with four-fermion interaction 
\cite{Chernodub2,KKZ}. Here, based on the HF approach, we present arguments in favor of the fact that not only in 
the massless but also in the {\it massive} (2+1)-D Thirring model (with nonzero bare Haldane mass) a non-Hermitian 
ground state can arise.

\section{Summary and conclusions}

In the present paper we have studied the possibility of the  dynamical appearance of both 
Hermitian and non-Hermitian mass terms in the originally Hermitian (2+1)-dimensional Thirring 
model. The last possibility means that non-Hermiticity can appear spontaneously in the model.

First of all, we consider dynamical symmetry breaking and fermion mass generation in the {\it massless} version 
(\ref{t1}) of the Thirring model. As it is shown in section \ref{IIA}, in this case the model is invariant under 
transformations from $U(2N)$ group, which contains two continuous chiral subgroups, $U(1)_{\gamma^5}$ and 
$U(1)_{\gamma^3}$ (\ref{n4}). Moreover, it is also symmetric with respect to several discrete transformations, 
two space reflections (or parity), $\cP_3$ and $\cP_5$, and  two time reversals, $\cT_3$ and $\cT_5$.
As a consequence, the massless (2+1)-D Thirring model is invariant under four different 
discrete $\cP_k\cT_l$ (where $k,l=3,5$) transformations (see in the Table I). 

The problem of dynamical mass generation is investigated using a nonperturbative HF approach based on the CJT 
effective action $\Gamma(S)$ (\ref{0360}) for the composite bifermion operator $\overline\Psi_k (x)\Psi_k (y)$. 
In fact, $\Gamma(S)$ is a functional of a full fermion propagator $S(x,y)$ (see in the section \ref{IIB}). In this 
case, in order to find the true fermion propagator of the initially massless Thirring model and to determine what 
kind of fermionic mass terms, Hermitian ${\cal M}_{H}$ (\ref{t13}) or non-Hermitian ${\cal M}_{NH}$ (\ref{t0140}), 
can arise dynamically in the model, it is sufficient to consider both the stationary equation (\ref{0370}) and
the CJT effective action (\ref{0360}) itself up to a first order in the coupling constant $G$ (see Eqs. (\ref{n043}) 
and (\ref{n420}), respectively). This is the essence of the Hartree-Fock method, which was used previously to prove 
the possibility of spontaneous non-Hermiticity in the massless (2+1)-D GN model \cite{KKZ}.  

Using HF approach, we have shown that due to the behavior (\ref{t75}) of the bare coupling constant $G(\Lambda)$ vs 
cutoff parameter $\Lambda$, the gap HF equation  (\ref{n043}) can be renormalized, i.e. reduced to a form which does 
not contain bare coupling $G$, but instead it depends only on the renormalization group invariant and finite free 
model parameter $g$ (it is introduced by Eq. (\ref{t75})).  More significant is the fact that this renormalized HF equation has 
two sets of solutions. Conventionally, they can be called as a set of (i) Hermitian and (ii) non-Hermitian solutions.  

At $g<0$, each solution from the Hermitian set (i) corresponds to the dynamical appearance of one or another 
Hermitian mass term ${\cal M}_{H}$ of the form (\ref{t13}) in the model. In this case, a spontaneous violation of 
one or another, discrete or continuous, symmetry of the model occurs, and qualitatively different phases
can arise spontaneously in the massless (2+1)-D Thirring model. Quasiparticle excitations of their ground states 
have the same mass $M_F$ equal to $|g|$, but the dynamics of these excitations is described by 
Lagrangians with different {\it Hermitian} mass terms (\ref{t13}). All these Hermitian phases of the model are 
degenerated, because the free energy density of their ground states is the same, and it is equal to $\frac{1}{48\pi}g^3$. 

In contrast, at $g>0$ the set (i) consists of a single solution with $m_H=m_D=m_3=m_5=0$. In this case there are no
symmetry breaking in the model. Quasiparticle excitations of its ground state are massless and their dynamics is 
described by the initially Hermitian massless Thirring Lagrangian (\ref{t1}).

Furthermore, we have shown that in the massless (2+1)-D Thirring model, each solution $\overline{S}(p)$ of the HF 
equation (\ref{n043}) from a non-Hermitian set (ii) corresponds to a dynamical appearance of a non-Hermitian mass term 
${\cal M}_{NH}$ of one of the forms (\ref{472}), (\ref{t20})-(\ref{t22}). As a result, in this case both at $g<0$ and 
$g>0$ the model can implement such phases in which quasiparticle excitations of their ground states have a real mass 
spectrum, but, however, the dynamics of these excitations is described by Lagrangians with non-Hermitian mass terms, 
i.e. spontaneous generation of non-Hermiticity occurs. All these non-Hermitian phases of the massless
(2+1)-D Thirring model are qualitatively different in the sense that their ground states differ significantly 
from each other in terms of their symmetry properties. Indeed, several non-Hermitian phases can be realized in the 
model, in which one or another $\cP_k\cT_l$ (where $k,l=3,5$) symmetry remains unbroken. However, there are also such 
non-Hermitian phases in which all $\cP_k\cT_l$ symmetries are spontaneously broken, etc (for details, see
in Sec. IV). 

It is interesting to note that at each fixed value of $g$ the variety of all Hermitian and non-Hermitian phases is 
degenerated, i.e. the ground states of all these phases have the same free energy density, and the phases, both
Hermitian and non-Hermitiian, can appear spontaneously in the {\it massless} (2+1)-D Thirring model (\ref{t1}) 
on the same footing. It means that the genuine vacuum (or ground state) of the (2+1)-D Thirring model is a mixed 
phase (or state). It can be imagined as a space, filled with one of these degenerated phases, in which bubbles of 
another, Hermitian and/or non-Hermitian, phases can be created.

Finally, in section V, we have shown that spatially homogeneous spontaneous non-Hermiticity can also be realized in the 
{\it massive} (2+1)-D Thirring model, but only when the bare Haldane mass term is nonzero. (In contrast, 
spontaneous non-Hermiticity was not found in the massive (2+1)-D GN model \cite{KKZ} as well as in the massive NJL
model \cite{Chernodub2}.) It turns out that 
in this case (and at arbitrary signs of $g$) the true vacuum of the model is indeed the mixed phase composed of a 
single Hermitian phase, in which the dynamics of quasiparticles is described by a Lagrangian with a Hermitian mass 
term of the form (\ref{t13}) with $m_D=m_3=m_5=0$, $m_H=(m_H)_0$ (\ref{t752}), and also of some non-Hermitian phases. 
The latter are described by Lagrangians with non-Hermitian mass terms (for details, see at the end of Sec. \ref{VA2}).   

Notice that in the framework of the HF approach, the effect of spontaneous non-Hermiticity of the model under 
consideration can be detected only at finite $N$, i.e. outside the leading order of the large-$N$ expansion technique.
And just the Fock term of the HF equation (\ref{n043}) plays the basic role in its appearance both in the 
(2+1)-D Thirring and GN \cite{KKZ} models. In addition, we are sure that using the HF method it is possible to show 
that in the generalized (2+1)-D Thirring model previously considered in \cite{KGK2}, a non-Hermitian ground state 
can also arise spontaneously.

We hope that the results of this article can be useful for describing physical phenomena in 
condensed matter systems having a planar crystal structure, or in thin films, e.g., like graphene. 
In such situations, it often happens that the elementary excitations of the system are massless. 
As a result, at low energies and in the continuum limit, their physical phenomena can be effectively 
described by {\it massless} quantum field theory models with four-fermion interactions of the 
type (\ref{t1}) \cite{Gusynin,Ebert,Mesterhazy}. And just in these cases, the effect of spontaneous 
non-Hermiticity could be manifested.

\section{ACKNOWLEDGMENTS}

R.N.Z. is grateful for support of the Foundation for the Advancement of Theoretical Physics and 
Mathematics BASIS.

\appendix
\section{Algebra of the $\gamma$ matrices in the case of $SO(2,1)$ group}
\label{ApC}

The two-dimensional irreducible representation of the (2+1)-dimensional
Lorentz group $SO(2,1)$ is realized by the following $2\times 2$
$\tilde\gamma$-matrices:
\begin{eqnarray}
\tilde\gamma^0=\sigma_3=
\left (\begin{array}{cc}
1 & 0\\
0 &-1
\end{array}\right ),\,\,
\tilde\gamma^1=i\sigma_1=
\left (\begin{array}{cc}
0 & i\\
i &0
\end{array}\right ),\,\,
\tilde\gamma^2=i\sigma_2=
\left (\begin{array}{cc}
0 & 1\\
-1 &0
\end{array}\right ),
\label{C1}
\end{eqnarray}
acting on two-component Dirac spinors $\psi(x)$. They have the properties:
\begin{eqnarray}
Tr(\tilde\gamma^{\mu}\tilde\gamma^{\nu})=2g^{\mu\nu};~~
[\tilde\gamma^{\mu},\tilde\gamma^{\nu}]=-2i\varepsilon^{\mu\nu\alpha}
\tilde\gamma_{\alpha};~
~\tilde\gamma^{\mu}\tilde\gamma^{\nu}=-i\varepsilon^{\mu\nu\alpha}
\tilde\gamma_{\alpha}+g^{\mu\nu},
\label{C2}
\end{eqnarray}
where $g^{\mu\nu}=g_{\mu\nu}=diag(1,-1,-1),
~\tilde\gamma_{\alpha}=g_{\alpha\beta}\tilde\gamma^{\beta},~
\varepsilon^{012}=1$.
There is also the relation:
\begin{eqnarray}
Tr(\tilde\gamma^{\mu}\tilde\gamma^{\nu}\tilde\gamma^{\alpha})=
-2i\varepsilon^{\mu\nu\alpha}.
\label{C3}
\end{eqnarray}
Note that the definition of chiral symmetry is slightly unusual in
(2+1)-dimensions (spin is here a pseudoscalar rather than a (axial)
vector). The formal reason is simply that there exists no other $2\times 2$ matrix anticommuting with the Dirac matrices $\tilde\gamma^{\nu}$
which would allow the introduction of a $\gamma^5$-matrix in the
irreducible representation. The important concept of 'chiral'
symmetries  and their breakdown by mass terms can nevertheless be
realized also in the framework of (2+1)-dimensional quantum field
theories by considering a four-component reducible representation
for Dirac fields. In this case the Dirac spinors $\Psi(x)$ have the
following form:
\begin{eqnarray}
\Psi(x)=
\left (\begin{array}{cc}
\psi_{1}(x)\\
\psi_{2}(x)
\end{array}\right ),
\label{C4}
\end{eqnarray}
with $\psi_1,\psi_2$ being two-component spinors.
In the reducible four-dimensional spinor representation one deals
with 4$\times$4 $\gamma$-matrices:
$\gamma^\mu=diag(\tilde\gamma^\mu,-\tilde\gamma^\mu)$, where
$\tilde\gamma^\mu$ are given in (\ref{C1}) (This particular reducible representation for 
$\gamma$-matrices is used, e.g., in Ref. \cite{Appelquist}). One can easily show, that
($\mu,\nu=0,1,2$):
\begin{eqnarray}
&&Tr(\gamma^\mu\gamma^\nu)=4g^{\mu\nu};~~
\gamma^\mu\gamma^\nu=\sigma^{\mu\nu}+g^{\mu\nu};~~\nonumber\\
&&\sigma^{\mu\nu}=\frac{1}{2}[\gamma^\mu,\gamma^\nu]
=diag(-i\varepsilon^{\mu\nu\alpha}\tilde\gamma_\alpha,
-i\varepsilon^{\mu\nu\alpha}\tilde\gamma_\alpha).
\label{C5}
\end{eqnarray}
In addition to the  Dirac matrices $\gamma^\mu~~(\mu=0,1,2)$ there
exist two other matrices, $\gamma^3$ and $\gamma^5$, which anticommute
with all $\gamma^\mu~~(\mu=0,1,2)$ and with themselves
\begin{eqnarray}
\gamma^3=
\left (\begin{array}{cc}
0~,& I\\
I~,& 0
\end{array}\right ),\,
\gamma^5=\gamma^0\gamma^1\gamma^2\gamma^3=
i\left (\begin{array}{cc}
0~,& -I\\
I~,& 0
\end{array}\right ),\,\,\tau=-i\gamma^3\gamma^5=
\left (\begin{array}{cc}
I~,& 0\\
0~,& -I
\end{array}\right )
\label{C6}
\end{eqnarray}
with  $I$ being the unit $2\times 2$ matrix.

\section{HF approach to the CJT effective potential $V(S)$}
\label{ApB}

\subsection{The case of massless Thirring model}\label{ApB1} 

In this case we have $D_\alpha^\beta(x,y)=
\left(\gamma^\nu\right)_\alpha^\beta i\partial_\nu\delta^3(x-y)$, so $\overline D(p)=\hat p\equiv\gamma^\nu p_\nu$.
It is clear from definition (\ref{t76}) that in order to get an effective potential $V(S)$ in the HF approximation, 
it is necessary to calculate the CJT effective action (\ref{n420}) using there for the full fermion propagator $S(x,y)$ the 
expression presented by Fourier transformation $\overline{S}(p)$ (\ref{t140})-(\ref{t14}). In addition, in this case  
the bare coupling $G$ is defined by its asymptotic behavior (\ref{t75}). As a result, the CJT effective action 
$\Gamma(S)$ (\ref{n420}) looks like
\begin{eqnarray}
\Gamma(S)=\Gamma_1+\Gamma_2+\Gamma_3+\Gamma_4,\label{B1}
\end{eqnarray}
where 
\begin{eqnarray}
\Gamma_1&\equiv&-i{\rm Tr}\ln \big (-iS^{-1}\big )=-i\int d^3x\int\frac{d^3p}{(2\pi)^3}{\rm tr}\ln\Big (
-i\overline{S^{-1}}(p)\Big )\nonumber\\ 
&=&-i\int d^3x\int\frac{d^3p}{(2\pi)^3}\ln\Det\big (\hat p+m_D+m_H\tau+im_5\gamma^5+im_3\gamma^3\big ), \label{B2}\\
\Gamma_2&\equiv& \int d^3xd^3y S^\alpha_\beta(x,y)D_\alpha^\beta(y,x)=
\int d^3x\int\frac{d^3p}{(2\pi)^3}
{\rm tr}\Big [\overline{S}(p)\overline D(p)\Big ]=\int d^3x\int\frac{d^3p}{(2\pi)^3}
{\rm tr}\Big [\overline{S}(p)\hat p\Big ]\nonumber\\
&=&i\int d^3x\int\frac{d^3p}{(2\pi)^3}\frac{4p^2(\Sigma^2+m_H^2-p^2)}
{\big ((\Sigma+m_H)^2-p^2\big )\big ((\Sigma-m_H)^2-p^2\big )},\label{B3}\\
\Gamma_3&\equiv&-\frac{G}2\int d^3x~ {\rm tr}\big [\gamma^\rho S(x,x)
\big ] {\rm tr}\big [\gamma_\rho S(x,x)\big ]=-\frac{G}2\int d^3x\int\frac{d^3p}{(2\pi)^3}
{\rm tr}\Big [\gamma^\rho\overline{S}(p)\Big ]\int\frac{d^3q}{(2\pi)^3}
{\rm tr}\Big [\gamma_\rho\overline{S}(q)\Big ],\label{B4}\\
\Gamma_4&\equiv&\frac{G}{2N}\int d^3x~ {\rm tr}\Big [\gamma^\rho S(x,x)\gamma_\rho S(x,x)\Big ]=
\frac{G}{2N}\int d^3x\int\frac{d^3p}{(2\pi)^3}\int\frac{d^3q}{(2\pi)^3}{\rm tr}\Big [\gamma^\rho\overline{S}(p)
\gamma_\rho\overline{S}(q)\Big ].\label{B5}\end{eqnarray}
Note that in Eq. (\ref{B2}) we have used for $\overline{S^{-1}}(p)$ the expression (\ref{t140}), as well as a rather 
general relation ${\rm tr}\ln A=\ln\Det A$. Moreover, in Eq. (\ref{B3}) we took into acount that $\hat p$ is indeed a 
Fourier image of the operator $D(x,y)$. Using Eq. (\ref{t14}), it is possible to show that (see the notations given in
Eq. (\ref{t16}))
\begin{eqnarray}
{\rm tr}\Big [\gamma^\rho\overline{S}(p)\Big ]=-\frac{4p^\rho}{\det(p)}\big (\Sigma^2+m_H^2-p^2\big ), \label{B6}
\end{eqnarray}
i.e. it is an odd expression vs each momentum $p^\rho$. As a result, in Eq. (\ref{B4}) each of the integrals on  
three-dimensional momenta is zero, and the entire expression $\Gamma_3$ is also zero. The determinant in  
Eq. (\ref{B2}) can be easily calculated, so we have
\begin{eqnarray}
&&\Gamma_1=-i\int d^3x\int\frac{d^3p}{(2\pi)^3}\ln\big 
[\big ((\Sigma+m_H)^2-p^2\big )\big ((\Sigma-m_H)^2-p^2\big )\big ].\label{B7}
\end{eqnarray}
Performing in the integrals of Eqs. (\ref{B3}) and (\ref{B7}) a Wick rotation, $p_0\to i p_3$, and then using in 
the obtained 
three-dimensional Euclidean integration space the spherical coordinate system, $p_3=p\cos\theta, 
p_1=p\sin\theta\cos\phi, p_2=p\sin\theta\sin\phi$, we have (after integration over angles, $0\le\theta\le\pi, 
0\le\phi\le2\pi$, and cutting off the region of integration of the variable $p$, $0\le p\le \Lambda$) 
the following asymptotic expansions for $\Gamma_1$ and $\Gamma_2$ at large values of $\Lambda$
\begin{eqnarray}
\Gamma_1&=&\frac{1}{2\pi^2}\int d^3x\Big\{2\Lambda (\Sigma^2+m_H^2)-\frac{\pi}{3}|\Sigma+m_H|^3-
\frac{\pi}{3}|\Sigma-m_H|^3+\cdots\Big\},  \label{B8}\\
\Gamma_2&=&\frac{1}{2\pi^2}\int d^3x\Big\{-4\Lambda (\Sigma^2+m_H^2)+\pi |\Sigma+m_H|^3+
\pi |\Sigma-m_H|^3+\cdots\Big\}, \label{B9}
\end{eqnarray}
where three dots mean the terms which disappear at $\Lambda\to\infty$.

It is obvious that only the first term, even over each of the momentum $p^\nu$, in expression (\ref{t14}) for the 
propagator $\overline{S}(p)$ will contribute to $\Gamma_4$. Then, taking into account that $\gamma^\rho=
diag(\tilde\gamma^\rho,-\tilde\gamma^\rho)$ (see in Appendix \ref{ApC}), one can obtain from Eq. (\ref{t14})
\begin{eqnarray}
\Gamma_4&=&
\frac{-6G}{2N}\int d^3x\int\frac{d^3p}{(2\pi)^3}\int\frac{d^3q}{(2\pi)^3}\frac{1}{\det(p)\det(q)}\Big\{
a(p)a(q)-b(p)\bar b(q)-\bar b(p)b(q)+\bar a(p)\bar a(q)\Big\}.\label{B10}
\end{eqnarray}
Performing in the momentum integrals of Eq. (\ref{B10}) a Wick rotation into the Euclidean momentum space and 
making there the same operations that we did with analogous situation in the expressions (\ref{B7}) and (\ref{B3})
for $\Gamma_1$ and $\Gamma_2$, we obtain
\begin{eqnarray}\Gamma_4&=&
\frac{3G}{4\pi^4N}\int d^3x\left\{\left [\int_0^\Lambda p^2dp\frac{(\Sigma^2-m_H^2)(m_D-m_H)+p^2(m_D+m_H)}
{\big ((\Sigma+m_H)^2+p^2\big )\big ((\Sigma-m_H)^2+p^2\big )}\right ]^2\right.\nonumber\\
&&~~~~~~~+\left [\int_0^\Lambda p^2dp\frac{(\Sigma^2-m_H^2)(m_D+m_H)+p^2(m_D-m_H)}
{\big ((\Sigma+m_H)^2+p^2\big )\big ((\Sigma-m_H)^2+p^2\big )}\right ]^2\nonumber\\
&&~\left.+2(m_5^2+m_3^2)\left [\int_0^\Lambda p^2dp\frac{\Sigma^2-m_H^2+p^2}
{\big ((\Sigma+m_H)^2+p^2\big )\big ((\Sigma-m_H)^2+p^2\big )}\right ]^2\right\}.\label{B11}
\end{eqnarray}
Using in each square brackets of Eq. (\ref{B11}) the following general asymptotic expansion formula 
(at $\Lambda\to\infty$)
\begin{eqnarray}
\int_0^\Lambda x^2dx\frac{A+Bx^2}
{(x^2+m^2)(x^2+n^2)}=B\Lambda+\frac{\pi B(n^3-m^3)+\pi A(m-n)}{2(m^2-n^2)}+
\frac{B(n^2+m^2)-A}{\Lambda}+\cdots\label{B12}
\end{eqnarray}
where $m=|\Sigma+m_H|$ and $n=|\Sigma-m_H|$, as well as taking into account the expansion (\ref{t75}) for bare coupling constant $G$, 
we have
\begin{eqnarray}
\Gamma_4&=&
\frac{1}{2\pi^2}\int d^3x\Big\{2\Lambda (\Sigma^2+m_H^2)-\frac{\pi g}{2}(\Sigma^2+m_H^2)
-\pi |\Sigma+m_H|^3-
\pi |\Sigma-m_H|^3+\cdots\Big\}.\label{B13}
\end{eqnarray}
As a result, it follows from Eqs. (\ref{B8}), (\ref{B9}) and (\ref{B13}) as well as from the relation 
$\Gamma_3\equiv 0$ that if fermion propagator $S$ has the form (\ref{t14}), then HF effective action $\Gamma (S)$ (\ref{n420}) in 
the limit $\Lambda\to\infty$ looks like
\begin{eqnarray}
\Gamma(S)=\Gamma_1+\Gamma_2+\Gamma_3+\Gamma_4=-\frac{1}{12\pi}\int d^3x\Big\{3g(\Sigma^2+m_H^2)+2 |\Sigma+m_H|^3+
2 |\Sigma-m_H|^3\Big\},\label{B14}
\end{eqnarray}
and the corresponding effective potential $V(S)$ takes a form (\ref{t77}).

\subsection{HF approach to $V(S)$ in the case of nonzero bare mass terms}\label{ApB2}

{\bf The presence of nonzero bare Haldane mass.} In this case, to find the effective potential $V_H(S)$ 
in the HF approximation (here and below the subscript $H$ means that the model contains a nonzero bare 
Haldane mass (\ref{m01})), we start from the expression (\ref{B1})
for CJT effective action $\Gamma(S)$ in the massless case, in which for $\overline D(p)$ it is necessary to use 
$\overline D(p)=\hat p+\tau m_{0H}$ (see in the section V). Hence the CJT effective action $\Gamma_H(S)$ for the 
corresponding massive Thirring model has the form
\begin{eqnarray}
\Gamma_H(S)&=&\Gamma(S)+m_{0H}\int d^3x\int\frac{d^3p}{(2\pi)^3}{\rm tr}\Big [\overline{S}(p)\tau\Big ],
\label{B15}
\end{eqnarray}
where $\Gamma(S)$ is presented by Eq. (\ref{B14}). Only the part of expression (\ref{t14}) for $\overline{S}(p)$ 
that is even over each momentum $p^\nu$ will give a nonzero contribution to Eq. (\ref{B15}). So, performing in this 
expression a Wick rotation and introducing a cutoff parameter $\Lambda$ (for more details, see the text after Eq. 
(\ref{t18})), we have 
\begin{eqnarray}
\Gamma_H(S)&=&\Gamma(S)+\frac{m_{0H}}{2\pi^2}\int d^3x\int_0^\Lambda p^2dp\frac{4p^2m_H-4(\Sigma^2-m_H^2)m_H}
{\big ((\Sigma+m_H)^2+p^2\big )\big ((\Sigma-m_H)^2+p^2\big )}.
\label{B16}
\end{eqnarray}
Now, applying the asymptotic expansion (\ref{B12}) to the integral over $p$ in Eq.  (\ref{B16}), we obtain at 
$\Lambda\to\infty$
\begin{eqnarray}
\Gamma_H(S)&=&\Gamma(S)+\frac{m_{0H}}{2\pi^2}\int d^3x\left\{4m_H\Lambda+m_H^2{\cal O}\left (\frac{m_H}\Lambda\right )
\right\}.\label{B17}
\end{eqnarray}
It follows from asymptotic expansions (\ref{t75}) and (\ref{m041}) for bare quantities $G$ and $m_{0H}$, respectively,
that 
\begin{eqnarray}
m_{0H}=\pm\mu^2\frac{\pi}{4\Lambda}+\mu{\cal O}\left (\frac{\mu^2}{\Lambda^2}\right ). \label{B18}
\end{eqnarray}
Hence, taking into account this relation in Eq. (\ref{B17}), we have at $\Lambda\to\infty$ a finite and renormalization 
group invariant expression for the CJT effective action $\Gamma_H(S)$ in the HF approximation, 
\begin{eqnarray}
\Gamma_H(S)&=&\Gamma(S)\pm\frac{\mu^2 m_H}{2\pi}\int d^3x,
\label{B19}
\end{eqnarray}
where $\Gamma(S)$ is the HF approximation (\ref{B14}) of the CJT effective action of the massless Thirring model. 
Using the definition (\ref{t76}), it is possible to obtain from  Eq. (\ref{B19}) the corresponding 
effective potential $V_H^{mas}$ (\ref{B20}) of the massive Thirring model (\ref{m01}) with nonzero bare Haldane mass 
term (when the sign 'plus' is selected in Eqs. (\ref{B18}) and (\ref{B19}), for definitness).

{\bf The presence of nonzero bare Dirac mass.} In this case, to find the effective potential $V_D(S)$ in the HF 
approximation (here and below the subscript $D$ means that the model (\ref{t1}) is extended now by the bare 
Dirac mass term 
$m_{0D}\overline \Psi_k \Psi_k$), we start from the expression (\ref{B1})
for CJT effective action $\Gamma(S)$ in the massless case, in which for $\overline D(p)$ it is necessary to use 
$\overline D(p)=\hat p+m_{0D}$. Hence the CJT effective action $\Gamma_D(S)$, has the form
\begin{eqnarray}
\Gamma_D(S)&=&\Gamma(S)+m_{0D}\int d^3x\int\frac{d^3p}{(2\pi)^3}{\rm tr}\Big [\overline{S}(p)\Big ],\label{B020}
\end{eqnarray}
where $\Gamma(S)$ is presented again by Eq. (\ref{B14}). Only the part of expression (\ref{t14}) for $\overline{S}(p)$ 
that is even over each momentum $p^\nu$ will give a nonzero contribution to Eq. (\ref{B020}). So, performing in this 
expression a Wick rotation and introducing a cutoff parameter $\Lambda$ (for more details, see the text after Eq. 
(\ref{t18})), we have 
\begin{eqnarray}
\Gamma_D(S)&=&\Gamma(S)+\frac{m_{0D}}{2\pi^2}\int d^3x\int_0^\Lambda p^2dp\frac{4p^2m_D+4(\Sigma^2-m_H^2)m_D}
{\big ((\Sigma+m_H)^2+p^2\big )\big ((\Sigma-m_H)^2+p^2\big )}.\label{B21}
\end{eqnarray}
Due to the asymptotic expansion (\ref{B12}), we have from Eq. (\ref{B21})
\begin{eqnarray}
\Gamma_D(S)&=&\Gamma(S)+\frac{m_{0D}}{2\pi^2}\int d^3x\left\{4m_D\Lambda+m_D^2{\cal O}\left (\frac{m_D}\Lambda\right )
\right\}.
\label{B22}
\end{eqnarray}
Finally, we should take into account in Eq. (\ref{B22}) the asymptotic expansion 
\begin{eqnarray}
m_{0D}=\pm\mu^2\frac{\pi}{4\Lambda}+\mu{\cal O}\left (\frac{\mu^2}{\Lambda^2}\right ), \label{B23}
\end{eqnarray}
which follows from Eqs. (\ref{t75}) and (\ref{m042}).
As a result, at $\Lambda\to\infty$ we have from Eq. (\ref{B22}) a finite and renormalization group invariant 
expression for the CJT effective action $\Gamma_D(S)$ in the HF approximation, 
\begin{eqnarray}
\Gamma_D(S)&=&\Gamma(S)\pm\frac{\mu^2 m_D}{2\pi}\int d^3x,
\label{B24}
\end{eqnarray}
where $\Gamma(S)$ is the HF approximation (\ref{B14}) to the CJT effective action of the {\it massless} Thirring model. 
Using the definition (\ref{t76}), it is possible to obtain from  Eq. (\ref{B24}) the corresponding 
effective potential $V_D^{mas}$ (\ref{B202}) of the massive Thirring model with nonzero bare Dirac mass 
term (when the sign 'plus' is selected in Eq. (\ref{B23}), for definitness).

\end{document}